\documentclass[aps,twocolumn,showpacs,prb,longbibliography]{revtex4-2}
\usepackage{mathtools}
\usepackage{bm}
\usepackage{dsfont,amsthm,amsbsy}
\usepackage{verbatim}
\usepackage{amssymb}
\usepackage{amsmath}
\usepackage{bbm}
\usepackage{overpic}
\usepackage{graphicx}
\usepackage{epstopdf}
\usepackage{subfigure}
\usepackage{natbib}
\usepackage{epsfig}
\usepackage{amsfonts}
\usepackage{mathrsfs}
\usepackage{sidecap}
\usepackage{lipsum}
\usepackage{url}
\usepackage[toc,page,title,titletoc,header]{appendix}
\usepackage[colorlinks,linkcolor=blue,citecolor=blue,anchorcolor=blue,urlcolor=blue]{hyperref}
\usepackage{hyperref}
\usepackage{resizegather}
\usepackage{tikz}
\usepackage{float}
\usepackage{mathbbol}
\usepackage[normalem]{ulem}
\usepackage{cancel}
\usepackage{upgreek}
\usepackage[T1]{fontenc}
\usepackage{enumitem}
\usepackage{verbatim}

\newcommand{\bl}{\begin{aligned}}
\newcommand{\el}{\end{aligned}}

\def\be{\begin{equation}}
\def\ee{\end{equation}}

\def\bi{\begin{itemize}}
\def\ei{\end{itemize}}
\def\bn{\begin{enumerate}}
\def\en{\end{enumerate}}
\def\bea{\begin{eqnarray}}
\def\eea{\end{eqnarray}}

\def\ba{\begin{array}}
\def\ea{\end{array}}
\def\bd{\begin{displaymath}}
\def\ed{\end{displaymath}}

\def\ket#1{\left|#1\right\rangle}
\def\bra#1{\left\langle#1\right|}

\def\tr{{\rm tr}}

\begin{document}

\title{Crossover from Quantum Chaos to a Reversed Quantum Disentangled Liquid\\ in a Disorder-Free Spin Ladder}
\author{Hanieh Najafzadeh}
\email[]{honeynjf@gmail.com}
\author{Abdollah Langari}
\email[]{abdollah.langari@gmail.com}
\affiliation{Department of Physics, Sharif University of Technology, Tehran 1458889694, Iran}
\begin{abstract}
	The mechanisms by which isolated interacting quantum systems evade thermalization extend beyond disorder-induced many-body localization, encompassing a growing class of interaction-driven phenomena. We investigate a spin-$\tfrac{1}{2}$ ladder with asymmetric XY leg couplings and tunable Ising interactions on the rungs, and identify the microscopic origin of quasi many-body localization (quasi-MBL) in this setting. Through a suite of diagnostics—including entanglement dynamics, fidelity susceptibility, adiabatic gauge potential norms, level-spacing statistics and entropy of eigenstates—we uncover a reentrant progression of dynamical regimes as the rung coupling $J_z$ is varied: integrable behavior at $J_z=0$, quantum chaos at intermediate $J_z$, and a robust nonthermal regime at strong coupling. In the latter regime, we demonstrate the emergence of a \emph{reversed quantum disentangled liquid} (reversed-QDL), where the light species thermalizes while the heavy species remains localized. The strong-coupling limit further yields emergent local integrals of motion anchored in a fixed-point structure, providing a microscopic origin of the observed quasi-MBL dynamics. These results establish reversed-QDL as a distinct, disorder-free route to nonergodicity and broaden the classification of dynamical phases in quantum matter.
\end{abstract}

\maketitle

\section{Introduction}

The breakdown of thermalization in isolated quantum systems has emerged as a central topic in nonequilibrium many-body physics. While generic interacting systems are expected to thermalize according to the eigenstate thermalization hypothesis (ETH)~\cite{Deutsch1991,Srednicki1994,Rigol2008,DAlessio2016}, exceptions to this rule—such as many-body localization (MBL) in disordered systems~\cite{Basko2006,Oganesyan2007,Nandkishore2015,Sierant2025,Abanin2019,Buca2023}—have revealed the rich landscape of nonergodic quantum phases. More recently, attention has shifted toward disorder-free mechanisms of thermalization failure, including Hilbert space fragmentation~\cite{Sala2020,Khemani2020,Jeyaretnam2025,Yang2025}, quantum many-body scars~\cite{Turner2018,Serbyn2021,Srivatsa2023,Yarloo2024}, mixed-phase space \cite{Michailidis2020}, Stark-MBL \cite{Evert2019,Schulz2019} and quantum disentangled liquids (QDLs)~\cite{Grover2014}.

QDLs represent a class of nonthermal states in which subsystems with different dynamical scales decouple entanglement-wise: typically, a ``heavy'' species thermalizes while a ``light'' species remains localized. Originally proposed in systems with mass imbalance or internal constraints, QDLs provide a framework for understanding partial ergodicity in constrained quantum dynamics~\cite{Grover2014,Veness2017,Abbasgholinejad2023,Smith2017}. In parallel, quasi–many-body localization (quasi-MBL)—characterized by slow entanglement growth and long-lived memory retention in the absence of quenched disorder—has been observed in spin ladders and other constrained, disorder-free settings~\cite{Yao2016,Yarloo2018,Sala2024,Gunawardana2024,Kuno2020,Orito2021,Cheng2024}. We stress that by quasi-MBL we refer to disorder-free, finite-size localization-like dynamics, which are distinct from true MBL driven by randomness.

While Hilbert-space fragmentation and quantum many-body scars also represent disorder-free mechanisms of nonergodicity, their microscopic origins are fundamentally distinct from the scenario studied here. Fragmentation arises from exactly preserved subspaces that confine dynamics~\cite{Moudgalya2022}, while scars correspond to atypical eigenstates embedded in an otherwise thermal spectrum~\cite{Chandran2023}. By contrast, our mechanism is rooted in emergent local integrals of motion (LIOMs)~\cite{Ros2015,Serbyn2013} that originate from the strong-coupling fixed point, where the integrals of motion are determined by the rung configurations. At this fixed point, the dynamics are effectively frozen by the dominance of the inter-leg coupling, and the system admits a set of integrals of motion. For large but finite $J_z$, these LIOMs remain quasi-conserved, giving rise to slow dephasing and logarithmic entanglement growth characteristic of reversed-QDL dynamics. Moreover, the reversed-QDL  regime identified in this work reflects an inverted entanglement hierarchy between dynamical species, a feature absent in both fragmentation and scarred dynamics. Our results therefore establish a conceptually distinct and complementary route to disorder-free nonergodicity.

Despite these advances, the microscopic mechanism underlying quasi-MBL in such systems remains poorly understood. In this work, we address this question in a spin-\(\frac{1}{2}\) ladder model featuring XY interactions along the legs and tunable Ising couplings across the rungs (see Fig.~\ref{fig:ladder}). We perform a comprehensive numerical analysis using complementary diagnostics, including entanglement dynamics~\cite{Bardarson2012,Dumitrescu2017}, fidelity susceptibility~\cite{GU2010,Sels2021,kim2025definingclassicalquantumchaos,Swietek2025}, adiabatic gauge potential (AGP) norms~\cite{Kolodrubetz2017,Pandey2020,LeBlond2021,Kim2024}, and level-spacing statistics~\cite{Oganesyan2007,Atas2013,Santos2010}. In the presence of (effective) mass imbalance, we reveal a reentrant structure in the system’s thermalization behavior as a function of the rung coupling: the ladder evolves from an integrable, nonergodic regime to a chaotic ETH-consistent phase, and ultimately enters a robust nonthermal regime at strong coupling.

Remarkably, we identify the latter regime as a \emph{reversed quantum disentangled liquid}, in which the light species thermalize while the heavy ones remain localized—reversing the entanglement hierarchy observed in conventional QDLs. This behavior uncovers the microscopic mechanism for quasi-MBL in translation invariant systems and establishes a two-scenario framework for the emergence of nonergodic dynamics without disorder. Our findings broaden the landscape of thermalization failure in many-body systems and point toward a novel class of stable, interaction-induced nonergodic phases.

Previous works have shown that disorder is not strictly necessary to induce localization. Instead, interactions in translation-invariant systems—particularly in spin ladders—can give rise to quasi-MBL, characterized by anomalously slow entanglement growth and long-lived memory~\cite{Yao2016,Kuno2020,Orito2021,Gunawardana2024,Sala2024,Yarloo2018}. However, these studies primarily focused on phenomenological features and prethermal behavior, without uncovering the mechanism responsible for long-time localization-like dynamics.
In this work, we move beyond phenomenological descriptions and provide a systematic diagnostic of thermalization and localization dynamics in the ladder model. Our central goal is to uncover the physical mechanism underlying the quasi-MBL behavior observed in Fig.~\ref{fig:ent_dyn}, and to demonstrate how it arises purely from interaction asymmetry and coupling structure.

Section~\ref{Model} introduces the model, while Sec.~\ref{Entanglement_dynamics} outlines its dynamical properties. In Sec.~\ref{Diagnostics}, we present extensive numerical simulations that characterize the distinct regimes of the model. Section~\ref{Mechanism} is devoted to elucidating the mechanism of reversed-QDL behavior and its relation to the strong-coupling limit. Finally, Sec.~\ref{Summary} summarizes our findings and discusses their broader implications.

\section{Model}
\label{Model}
\begin{figure}[t]
	\centering
	\includegraphics[width=1.05\linewidth, trim=180 180 120 180, clip]{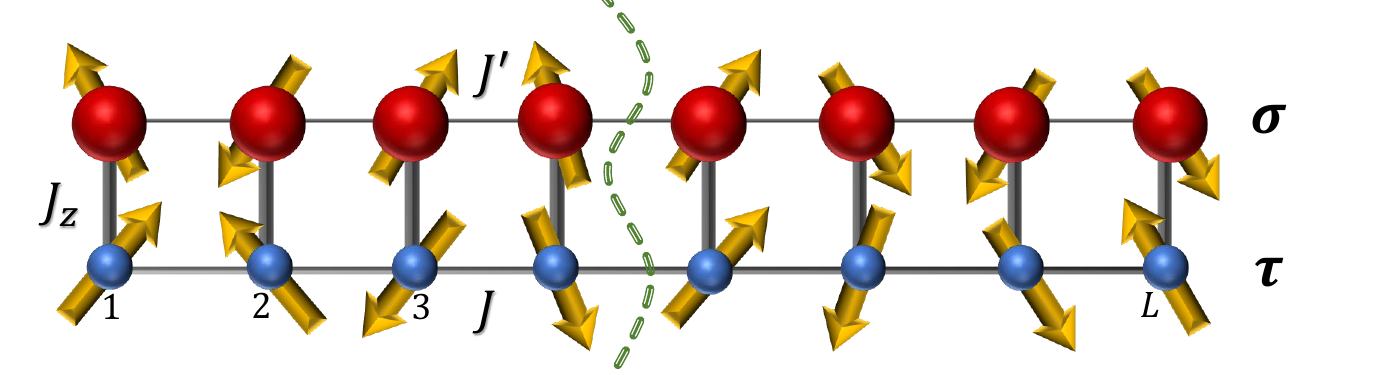}
\caption{%
	(Color online) Schematic illustration of the two-leg spin-\(\frac{1}{2}\) ladder model. Blue and red spheres denote spins on the lower (\(\tau\)) and upper (\(\sigma\)) legs, respectively, which are coupled via XY interactions \( J \) and \( J' \). Rung couplings \( J_z \) introduce inter-leg Ising interactions.
	The dashed green line denotes the bipartition used to measure entanglement entropy.
}
\label{fig:ladder}
\end{figure}

We consider a spin-\(\frac{1}{2}\) two-leg ladder composed of \( L \) rungs as shown in Fig.~\ref{fig:ladder}, where each rung consists of two spins corresponding to the lower (\(\tau\)) and upper (\(\sigma\)) ones, respectively~\cite{Yao2016}. The system is governed by the following Hamiltonian assuming periodic boundary conditions (PBC, i.e. $\sigma_{L+1} =
\sigma_1, \tau_{L+1} =
\tau_1$) along legs, $H=H_{\tau}+H_{\sigma}+H_{zz}$, where
\begin{eqnarray}
\label{eq:H}
H_{\tau} &=&  J \sum_{i=1}^{L} \left( \tau_{i}^{x} \tau_{i+1}^{x} + \tau_{i}^{y} \tau_{i+1}^{y} \right), \\
H_{\sigma} &=& J' \sum_{i=1}^{L} \left( \sigma_{i}^{x} \sigma_{i+1}^{x} + \sigma_{i}^{y} \sigma_{i+1}^{y} \right), \nonumber
\\
H_{zz} &=& J_z \sum_{i=1}^{L} \tau_{i}^{z} \sigma_{i}^{z}, \nonumber
\end{eqnarray}
where \( \tau_i^\alpha \) and \( \sigma_i^\alpha \) (\( \alpha = x, y, z \)) are Pauli operators acting on site \( i \) of the lower and upper legs, respectively. The coupling \( J_z \) introduces the strength of Ising interaction on each rung. The intra-leg XY interactions act with strength \( J \) on the lower leg and \( J' \) on the upper leg~\cite{Giamarchi2003}.

We fix \( J = 1 \) to define the energy scale, $J'=0.001$ and explore the physics as a function of the inter-leg coupling \( J_z \), focusing on the strongly asymmetric regime \( J' \ll J \). This limit induces a strong asymmetry in dynamical timescales between the two legs, enabling partial thermalization and nontrivial entanglement structures—key ingredients for QDL or quasi-MBL physics—, which is going to be uncovered in this article.

It would be interesting to mention that the XY Hamiltonian of each leg can be mapped to a spinless free fermion model via a Jordan-Wigner transformation~\cite{Jordan1928,Lieb1961}. Let us define $c_{\ell}^{\dagger}$ ($c_{\ell}$) as the 
creation (annihilation) of a fermion on site $\ell$ on the lower leg and the corresponding ones on the upper leg as $\tilde{c}_{\ell}^{\dagger}$ ($\tilde{c}_{\ell}$), where $\{c, \tilde{c}\}=0$. 
The details of transformation is given in  \hyperref[Jordan-Wigner]{Appendix~\ref*{Jordan-Wigner}}.
The Hamiltonian of the ladder will be written in terms of fermion operators as
\begin{eqnarray}
\label{Fermion_H}
 H_{\tau}&=&2 J \sum_{\ell=1}^{L} \left(c_{\ell}^{\dagger} c_{\ell+1} + c_{\ell+1}^{\dagger} c_{\ell}\right), \\
 H_{\sigma}&=&2 J' \sum_{\ell=1}^{L} \left(\tilde{c}_{\ell}^{\dagger} \tilde{c}_{\ell+1} + \tilde{c}_{\ell+1}^{\dagger} \tilde{c}_{\ell}\right), \nonumber \\
H_{zz}&=&4 J_z \sum_{\ell=1}^{L} n_{\ell} \tilde{n}_{\ell} - 2 J_z \sum_{\ell=1}^{L} \left(n_{\ell} + \tilde{n}_{\ell}\right) + const. \nonumber
\end{eqnarray}
The fermion representation of Hamiltonian (\ref{Fermion_H}) shows that the ladder is composed of two free fermion chains coupled via density-density interaction between aligned sites across each rung. It has been mentioned
in~\cite{Yao2016} that for $J'=0$, the density of fermions on the upper chain (or equivalently the spin configurations of $\sigma$'s) plays the role of a quenched disorder for the lower chain fermions, leading to (single particle) Anderson localization~\cite{Anderson1958} at strong disorder. Upon taking into account the effect of small $J'$
($J' \ll J$), the slow dynamics of heavy fermions ($\tilde{c}$) could act as slow variation of disorder landscape in an interacting fermion system. The long-time dynamics of the model shows a slow decay of polarization~\cite{Yao2016,Khemani2017}, which mimics an MBL-like dynamics dubbed quasi-MBL regime. The signature of the latter phase is decoded in the slow (logarithmic) growth of entanglement entropy as shown in Fig.~\ref{fig:ent_dyn} for large Ising coupling ($J_z=10, 20$). 
We would like to stress that all numerical simulations in this article are conducted on the spin Hamiltonian defined in Eq.~\eqref{eq:H}.
We are going to uncover the mechanism of slow dynamics and investigate the effect of Ising interaction on the dynamical behavior of the model within several diagnostics approaches.
\section{Entanglement dynamics}
\label{Entanglement_dynamics}

Entanglement entropy measures how strongly two parts of a quantum system are quantum-correlated, which is a universal probe of quantum structure, dynamics, and phases of matter~\cite{Amico2008,Laflorencie2016}. More specifically, the dynamical behavior of entanglement entropy gives a clear signature of the dynamical properties of the phase of system. In a thermal system (which obeys ETH or being in a chaotic phase), the entanglement entropy grows rapidly with time (typically ballistically, i.e. linear in time for 1D systems) and
eventually saturates to a volume law value~\cite{Rigol2008,DAlessio2016}. While in an MBL phase,
entanglement entropy grows logarithmically in time and saturates to a subthermal volume law (much smaller than in the thermal phase)~\cite{Bardarson2012}.

For a pure state, $|\psi(t)\rangle$, the density matrix is given by
$\rho(t)=|\psi(t)\rangle\langle\psi(t)|$ as the state evolves via the unitary evolution of the Hamiltonian (Eq.~\ref{eq:H}).
We compute the bipartite entanglement entropy,
\begin{equation}
 S_{\mathrm{ent}}(t) = -\mathrm{Tr}_A \rho_A \ln \rho_A,
 \label{S_ent}
\end{equation}
where \( \rho_A \) is the reduced density matrix obtained by tracing out half of the system across a central vertical cut (dashed line in Fig.~\ref{fig:ladder}),
i.e., $\rho_A=\mathrm{Tr}_B (\rho)$.
\begin{figure}[t]
	\centering
	\begin{overpic}[width=0.9\linewidth, trim=130 10 130 1, clip]{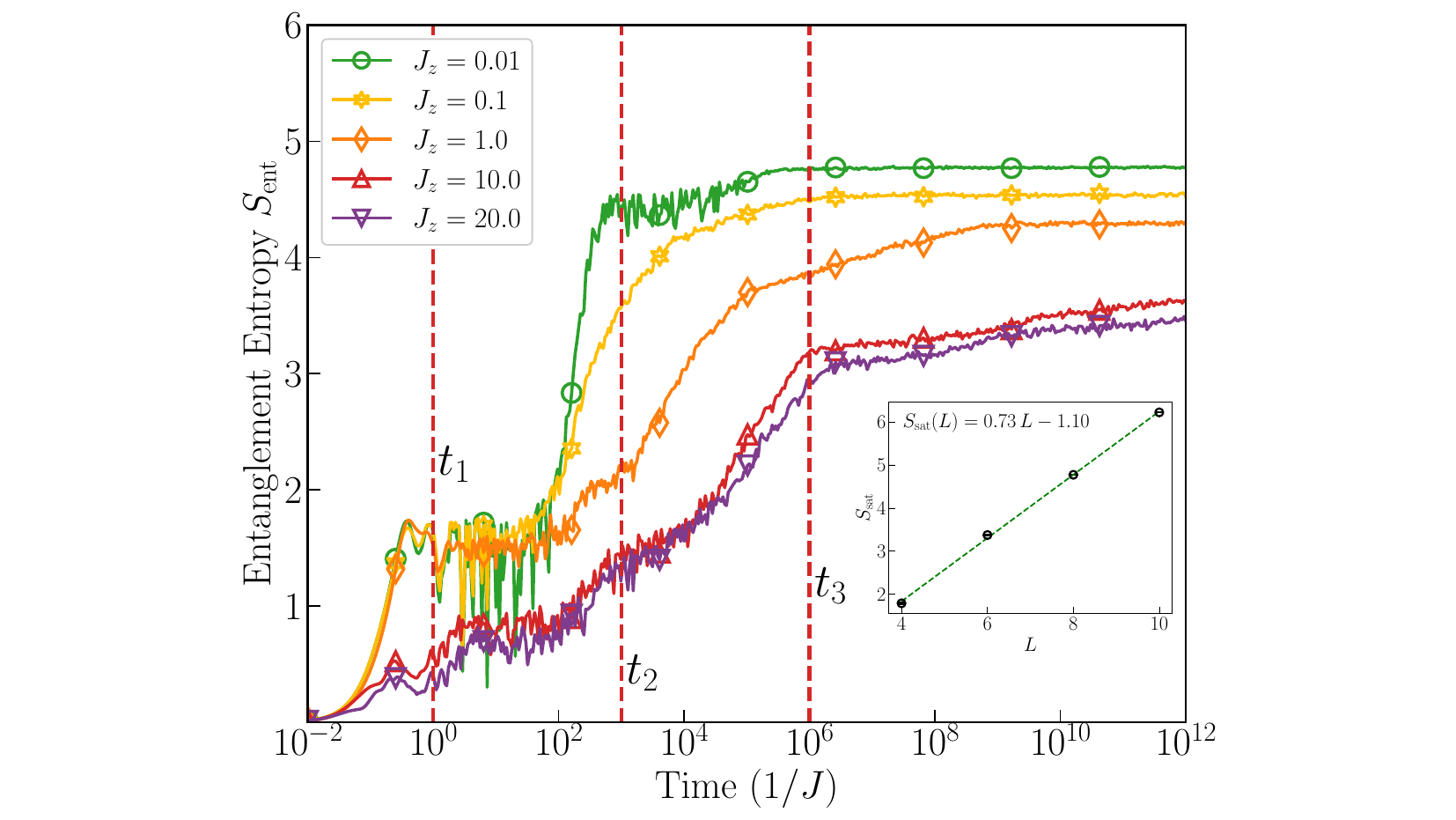}
		\put(76,23){\textcolor{black}{$J_z=0.01$}}
	\end{overpic}
	\caption{%
		(Color online) Time evolution of the half-cut entanglement entropy \( S_{\mathrm{ent}}(t) \) for various values of inter-leg Ising coupling \( J_z \), computed for a spin-1/2 ladder of size \( L = 8 \) with periodic boundary conditions and $J'=0.001$. The system exhibits a crossover from rapid thermalization at weak coupling (\( J_z \ll 1 \)) to slow, logarithmic growth characteristic of quasi-many-body localization (\( J_z \gg 1 \)). For \( J_z = 10 \), the labeled times \( t_1 \), \( t_2 \) and $t_3$ respectively mark the rapid rise due to local interactions, the onset and crossover to the logarithmic growth regime, highlighting the separation of dynamical timescales inherent to quasi-MBL.
		\textbf{(Inset)} The entropy saturation value plotted against system size at $J_z=0.01$
		demonstrates a clear volume-law scaling, supports the identification of
		a chaotic phase.
	}
	\label{fig:ent_dyn}
\end{figure}
The time evolution starts from an initial state, which is random product states sampled uniformly from the Hilbert space, and observables are averaged over 100 such realizations to suppress fluctuations. Time evolution is performed via exact diagonalization of the full Hamiltonian.
As shown in Fig.~\ref{fig:ent_dyn}, the dynamics of averaged half-cut entanglement entropy \( S_{\mathrm{ent}}(t) \) exhibits distinct behavior as a function of the inter-leg coupling \( J_z \), for $J=1, J'=0.001$. In the strongly interacting regime (\( J_z \gg 1 \)), \( S_{\mathrm{ent}}(t) \) shows a rapid initial rise due to
the dynamics of $\tau$ degress of freedom followed by a broad saturation-like plateau, which begins at $t_1 \sim (J)^{-1}$ and ends at $t_2 \sim (J')^{-1}$~\cite{Yao2016}.
Beyond this regime, the entanglement grows slowly and logarithmically
(with smaller slope)
in time—a hallmark of non-thermalizing phase in translation-invariant systems.
The latter regime ($t>t_3$) is responsible for a residual memory in a finite system before diminishing, however, the diminishing time scales with system size to infinity, according to the analysis presented in Ref.~\cite{Yao2016}.
The presence of this logarithmic regime at late times, which starts at $t_3$, reflects the emergence of a slow dephasing process across the ladder, driven solely by interaction-induced constraints.

In contrast, for weak inter-leg coupling (\( J_z \ll 1 \)), the entanglement dynamics grows fast and reaches its saturation value.
However, it reveals an additional plateau close to saturation at intermediate times, which ends before $t_3 \sim (J’^2/J_z)^{-1}$ \cite{Yao2016}. The appearance of this second plateau indicates that thermalization is enhanced within the fast leg—mediated by its interaction with the slow leg—before global dephasing occurs. This behavior signals more structured, multi-stage dynamics at small $J_z$.
Consistent with this picture, the inset of Fig.~\ref{fig:ent_dyn} shows that the saturation value at $J_z=0.01$ scales linearly with system size, i.e. it obeys a clear volume law. This scaling behavior provides direct evidence that the weak-coupling regime is chaotic and thermalizing, in sharp contrast to the logarithmic entanglement growth and sub-volume saturation characteristic of the quasi-MBL regime at large \( J_z \).

\section{Observables and Diagnostics}
\label{Diagnostics}

To characterize the 
emergent phases of the spin ladder model defined in Sec.~\ref{Entanglement_dynamics}, we employ a set of complementary observables 
to distinguish between thermal and nonthermal regimes across different coupling strengths \( J_z \).

\subsection{Fidelity susceptibility and adiabatic gauge potential norm}

The fidelity susceptibility (FS), defined as the diagonal component of the quantum geometric tensor with respect to a coupling parameter $\lambda$, provides a sensitive probe of the structure of eigenstates~\cite{You2007,GU2010}. For the $n$-th eigenstate $|\psi^{(n)}\rangle$, i.e. $H |\psi^{(n)}\rangle=E_n |\psi^{(n)}\rangle$, the fidelity susceptibility is expressed as
\begin{equation}
\chi_n \;=\; \sum_{m\neq n} \frac{|\langle \psi^{(n)}|\partial_\lambda H|\psi^{(m)}\rangle|^2}{(E_m - E_n)^2},
\label{Fidelity_susceptibility}
\end{equation}
quantifying the response of $|\psi^{(n)}\rangle$ to an infinitesimal change in $\lambda$.  Since the denominator involves squared energy differences, $\chi_n$ is exponentially sensitive to the fine structure of the many-body spectrum, making it a natural diagnostic of ergodic versus localized behavior~\cite{Pandey2020,Sels2021,kim2025definingclassicalquantumchaos,Swietek2025}.
Therefore, fidelity susceptibility is a valuable tool for illustrating how dynamical behavior changes between different regimes.

Since all eigenstates may contribute to the dynamics, it is natural to consider the fidelity susceptibility (FS) averaged over the spectrum. In order to capture the scaling behavior while reducing the effect of large eigenstate-to-eigenstate fluctuations, it is more convenient to analyze the typical value of $\chi_n$ obtained by averaging its logarithm~\cite{Pandey2020,Sels2021,LeBlond2021}. We thus define the \emph{typical fidelity susceptibility} as
\begin{equation}
\chi_{\mathrm{typ}}=e^{\zeta} \;;\;\; \zeta \;=\; \langle \ln(\chi_n) \rangle, 
\label{Zeta}
\end{equation}
where $\langle \cdots \rangle$ denotes averaging over different eigenstates,
 within each symmetry sector.
This log-average suppresses the influence of rare resonances while retaining the exponential sensitivity of the probe, making $\zeta$ particularly useful for distinguishing between ETH (chaotic), integrable, and MBL phases through their different scaling behaviors~\cite{LeBlond2021,Sels2023}.

\begin{figure}[t]
	\centering
\includegraphics[width=1.1\linewidth]{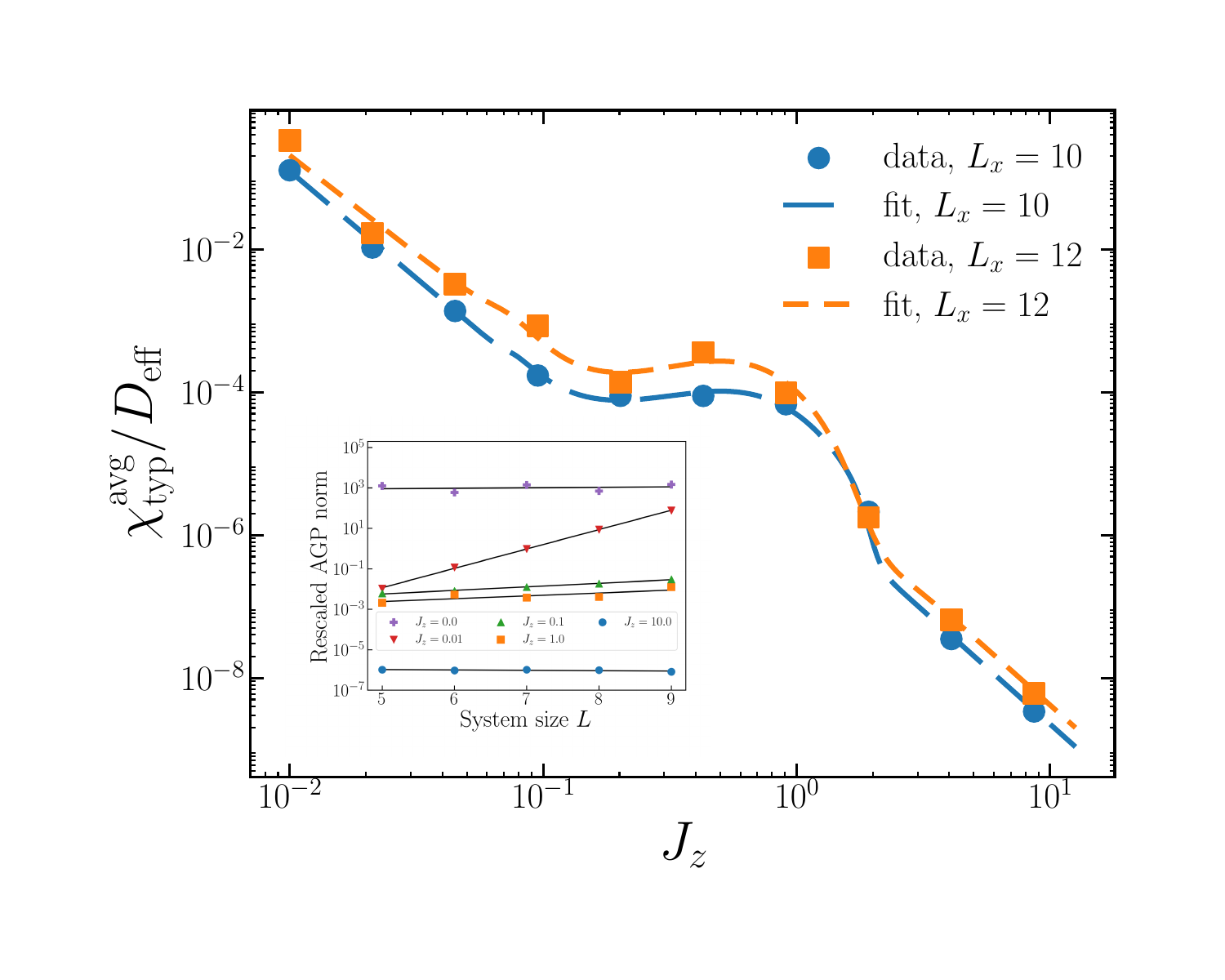}

\vspace{-0.2cm}
	\caption{
(Color online)
Dimension-weighted average of fidelity susceptibility 
$\chi_{\mathrm{typ}}^{\mathrm{avg}}/D_{\mathrm{eff}}$ as a function of $J_z$ for ladder lengths $L=10, 12$. A clear peak appears near $J_z \simeq 1$, signaling a dynamical crossover. 
(Inset) Rescaled AGP norm $\|A_\lambda\|^2 / L$ as a function of system size for different values of $J_z$ on a logarithmic scale. For $J_z=0$ (integrable case), the scaling of $\|A_\lambda\|^2$ is linear in $L$, while for small nonzero $J_z$ values the norm grows exponentially, with $J_z=0.01$ showing clear chaotic behavior. Near $J_z \simeq 1$, deviations from exponential scaling appear, and for larger $J_z$, the exponential growth is strongly reduced.
 The AGP results are computed in the total-magnetization sector $\mathcal{S}^z_{\mathrm{tot}}=0$ and averaged over all $(k,\mathcal{Z})$  symmetry sectors.
}
	\label{fig:fidelity_vs_Jz}
\end{figure}

In order to investigate the effect of the rung coupling $J_z$ on the dynamics of our model, we set the driving parameter to be $\lambda = J_z$. This choice leads to
\begin{equation}
\partial_\lambda H \equiv \frac{1}{L} \sum_{\ell=1}^L \tau^z_{\ell}\,\sigma^z_{\ell}\,
\label{Partial_H}
\end{equation}

\noindent which is the operator entering the calculation of $\chi_n$ in Eq.~\ref{Fidelity_susceptibility}. 
For each symmetry sector $\theta$ with dimension $D_{\theta}$, we compute the
	sector-resolved fidelity-susceptibility and form a dimension-weighted average
	\begin{equation}
	\chi_{\mathrm{typ}}^{\mathrm{avg}}(J_z)
	=\frac{\sum_{\theta} D_{\theta}\, \chi_{\mathrm{typ}}^{(\theta)}(J_z)}
	{\sum_\theta D_\theta}.
	\end{equation}
In Fig.~\ref{fig:fidelity_vs_Jz} we present, $\chi_{\mathrm{typ}}^{\mathrm{avg}}/D_{\mathrm{eff}}$, as a function of $J_z$ for different ladder lengths $L$,
where $D_{\mathrm{eff}}$ is the mean of the Hilbert space dimensions of all considered symmetry sectors.
A pronounced peak is observed near $J_z \simeq 1$ for $L=10, 12$, 
which provides clear evidence of a dynamical crossover between two regimes ~\cite{Pandey2020,Sels2021}.
Due to the strong finite-size effects, however, it is not possible to reliably extract a universal scaling function from the available data for the fidelity susceptibility.

All fidelity-susceptibility results are computed in the
symmetry-resolved sector $\theta:$ $(S^z_{\tau}, S^z_{\sigma}, k, \mathcal{Z})=(0,0,k_m,\pm 1)$ via the shift-and-invert method, focusing on the central region of the spectrum—capturing approximately 50\% of the states for $L=10$ and 10\% for $L=12$.
Here $S^z_{\tau}$ and $S^z_{\sigma}$ denote the total magnetizations on the two legs, 
$k_m=2\pi m/L$ is the lattice momentum and $\mathcal{Z}$ specifies the spin-inversion symmetry.
The data of Fig.~\ref{fig:fidelity_vs_Jz}(main) is averaged
over all non-reflection-symmetric momentum sectors ($k_m\neq 0, \pi$) for $L= 10$ and over
four generic momentum sectors for $L= 12$, namely $(0,0,\pi/3,\pm 1)$ and $(0,0,5\pi/6,\pm 1)$.

The adiabatic gauge potential (AGP) quantifies the generator of slow deformations in parameter space~\cite{Kolodrubetz2017,Sels_2017}. Its Hilbert--Schmidt (Frobenius) norm coincides with the eigenstate-averaged fidelity susceptibility~\cite{Pandey2020}. We evaluate the AGP with respect to the rung coupling $\lambda= J_z$ and use the regularized norm
\begin{equation}
	\|A_\lambda\|^2
	=\frac{1}{\mathcal{D}}\sum_{n}\sum_{m\neq n}
	\frac{\omega_{mn}^2\,\big|\langle\psi^{(m)}|\partial_\lambda H|\psi^{(n)}\rangle\big|^2}{\big(\omega_{nm}^2+\mu^2\big)^2},
	\label{AGP_norm}
\end{equation}
with $\omega_{mn}=E_m-E_n$, $\mathcal{D}$ the Hilbert-space dimension, and cutoff $\mu=L\,\mathcal{D}^{-1}$ chosen to suppress the zero-frequency divergence~\cite{Pandey2020,Kim2024}. 

\emph{Expected scaling.—} The AGP norm tracks many-body complexity~\cite{Pandey2020}. In ergodic/ETH regimes one expects entropy-controlled growth,
$\|A_\lambda\|^2\propto e^{S(L)}$,
with $S(L)\simeq sL$ and $s=\ln 4$ for our two-leg spin-$\tfrac12$ ladder. In the free noninteracting limit the AGP is quasi-local and extensive,
$\|A_\lambda\|^2\propto L$.
Generic integrable or weakly chaotic regimes exhibit sub-ETH behavior,
$\|A_\lambda\|^2\propto L^{\beta}$ ($\beta>1$) or
$\|A_\lambda\|^2\propto e^{\alpha S(L)}$ with $0<\alpha<1$.

The inset of Fig.~\ref{fig:fidelity_vs_Jz} shows the rescaled norm $\|A_\lambda\|^2/L$ versus $L$ on a logarithmic axis. Fitting
$\ln(\|A_\lambda\|^2/L)=\alpha(J_z)L+ const.$
over $L=5,...,9$ yields $\alpha=\{0.050,\,2.207,\,0.405,\,0.329,\,0.043\}$ for $J_z=\{0,\,10^{-2},\,10^{-1},\,1,\,10\}$, 
so that compactly
$\|A_\lambda\|^2/L\sim e^{\alpha(J_z)L}$.\\

These values align with the theoretical picture mentioned above. Unlike the FS, the AGP norm is a frequency-filtered and regularized spectral quantity, making it explicitly sensitive to how near-resonant level pairs contribute as the many-body level spacing decreases with increasing system size. As a result, the AGP norm naturally exhibits distinct size scaling for different values of \(J_z\), reflecting qualitative differences between ETH-like and non-ergodic regimes.

Distinguishing a mild exponential from a power law over $L=5$,...,$9$ is intrinsically difficult; nevertheless, the intermediate couplings display clear ETH-like exponential scaling tied to the entropy density $s=\ln 4$, whereas both integrable edges ($J_z=0$ and large $J_z$) remain strongly suppressed.


\subsection{Entropy of eigenstates}
The final quantity we compute to probe the dynamical behavior is the normalized average entropy~\cite{Bianchi2022},
\begin{equation}
	S_{\mathrm{avg}} = \frac{1}{\mathcal{D}} \sum_{n} \frac{S^{(n)}_A}{2 L_A \ln 2} \;,
	\label{S_avg}
\end{equation}
where $\mathcal{D}$ is the total Hilbert-space dimension and \( S^{(n)}_A \) denotes the half-cut entanglement entropy of the eigenstate \( |\psi^{(n)}\rangle \), as defined in Eq.~\eqref{S_ent}. Here \(L_A\) is the size (number of rungs) of the subsystem \(A\) used in the bipartition. For even \(L\) we set \(L_A=L/2\), while for odd \(L\) we take \(L_A=(L+1)/2\).

\begin{figure}[t]
	\centering
		\includegraphics[width=1.0\linewidth]{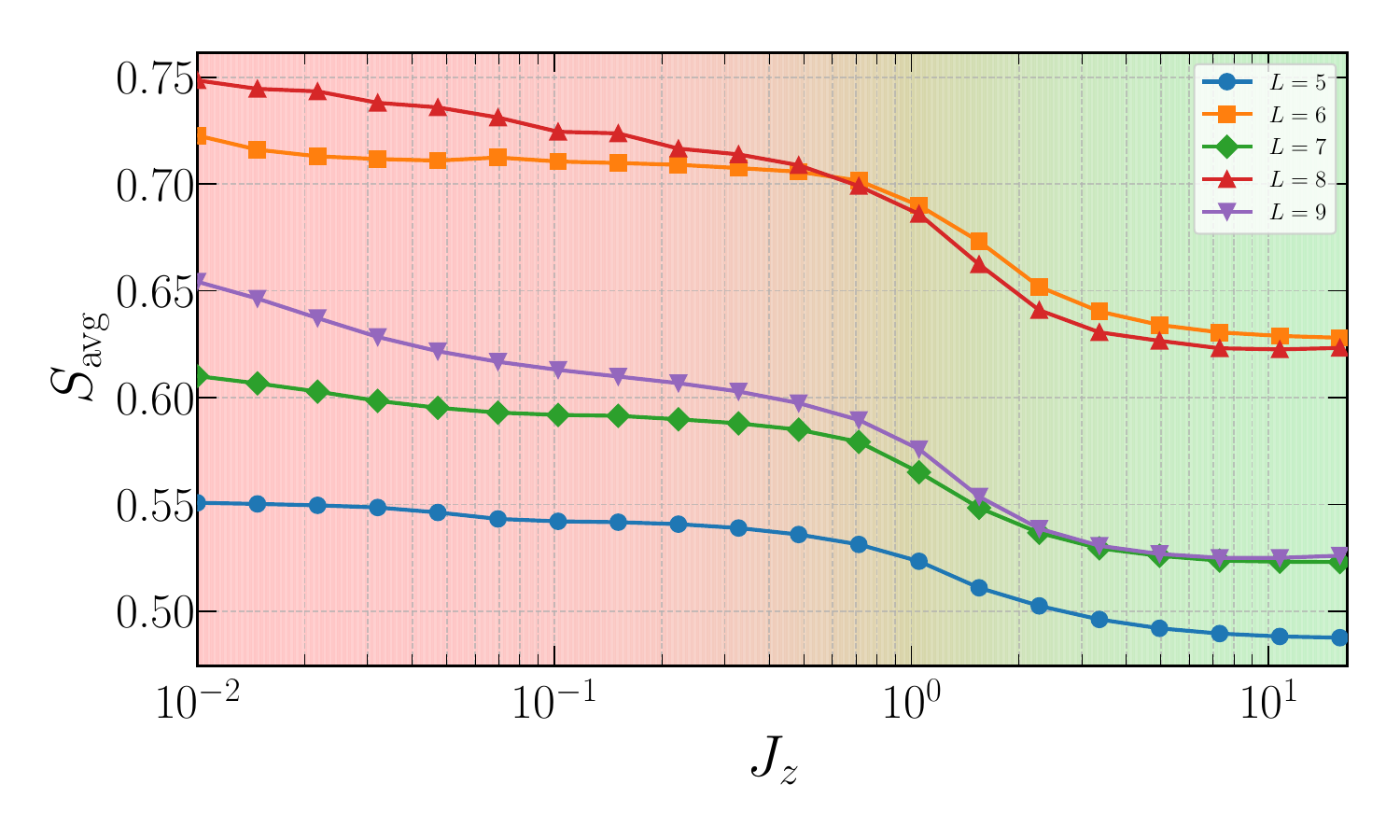}

	\caption{(Color online)  Normalized average entropy $S_{\mathrm{avg}}$ versus $J_z$ for system sizes $L=5,6,7,8,9$.
		For $J_z<1$, larger values of $S_{\mathrm{avg}}$ indicate full Hilbert-space exploration in the chaotic phase, while for $J_z>1$, a marked decrease signals localization, consistent with a dynamical crossover near $J_z\simeq 1$.
		Slight differences between odd and even $L$ arise from the bipartition asymmetry in entropy calculations.
	}
	\label{fig:average-entropy}
\end{figure}

We plot \( S_{\mathrm{avg}} \) over 80\% of the central spectrum as a function of \( J_z \) for system sizes \( L = 5, 6, 7, 8, 9 \) in Fig.~\ref{fig:average-entropy}.
The data exhibit two distinct regimes: for \( J_z < 1 \), \( S_{\mathrm{avg}} \) takes larger values, indicating full access to the Hilbert space consistent with a chaotic phase; for \( J_z > 1 \), \( S_{\mathrm{avg}} \) decreases clearly, signaling non-ergodicity in the Hilbert space. This behavior aligns with other numerical diagnostics
that is a supporting indicator for dynamical crossover around \( J_z \simeq 1 \).

It is also worth noting that results for odd and even system sizes differ slightly. This discrepancy arises from the bipartition cut used in calculating the entanglement entropy, which is symmetric for even \( L \) but asymmetric for odd \( L \).

We have also computed the mean level spacing as a function of $J_z$ 
(see \hyperref[app:level-spacing]{Appendix~\ref*{app:level-spacing}}), which demonstrates the crossover from Gaussian statistics at small $J_z$ to Poisson-type statistics for large $J_z$, consistent with the other diagnostics.

\section{Mechanism of Quantum Dynamics}
\label{Mechanism}

All quantum diagnostics discussed so far indicate that the spin ladder described by the Hamiltonian in Eq.~\ref{eq:H} exhibits three distinct dynamical phases/regimes as a function of \( J_z \). At \( J_z = 0 \), the model consists of two decoupled integrable spin-\(\tfrac{1}{2}\) chains. Introducing a small but finite \( J_z \) drives the system into a chaotic phase, which persists up to \( J_z \lesssim 1 \), where a dynamical crossover occurs. For \( J_z \gtrsim 1 \), the model exhibits quasi-MBL behavior, as summarized in the diagram shown in Fig.~\ref{fig:phase_diagram}.

Despite the absence of disorder, this translationally invariant system displays logarithmic entanglement growth, reminiscent of conventional MBL phases. Due to the strong asymmetry in the XY couplings along the two legs (\( J' \ll J \)), the system may also be interpreted as a quantum disentangled liquid (QDL): the heavier \(\sigma\)-spins thermalize rapidly and effectively localize the lighter \(\tau\)-spins, preventing their full thermalization. This interplay leads to the slow, logarithmic growth of entropy. In the following, we aim to uncover the underlying mechanism responsible for the quasi-MBL behavior observed at large \( J_z \).

\subsection{Measured Entanglement Entropy: QDL Diagnostic}

To probe the structure of partial thermalization, we compute the \emph{measured entanglement entropy}, a diagnostic originally introduced in the context of quantum disentangled liquids (QDLs)~\cite{Grover2014,Ben_Zion_2020,Smith2017}. In our setup, we treat the \(\sigma\) spins as the measured subsystem, and examine the entanglement properties of the \(\tau\) spins after projective measurements are performed on the \(\sigma\) spins.

	

Given an energy $E_m$ and its corresponding eigenstate \( |\psi^{(m)}\rangle \) of the Hamiltonian in Eq.~\eqref{eq:H}, we express it in the product basis of \(\tau\) and \(\sigma\) spins as
\begin{equation}
	|\psi^{(m)}\rangle = \sum_{i,j} C_{ij}^{(m)}\, |\mu_i^{(\tau)}\rangle \otimes |\eta_j^{(\sigma)}\rangle,
\end{equation}
where \( |\mu_i^{(\tau)}\rangle \) and \( |\eta_j^{(\sigma)}\rangle \) form complete basis sets for the \(\tau\) and \(\sigma\) spins, respectively.

\begin{figure}[t]
	\centering
	\includegraphics[width=1.\linewidth, trim=130 10 130 1, clip]{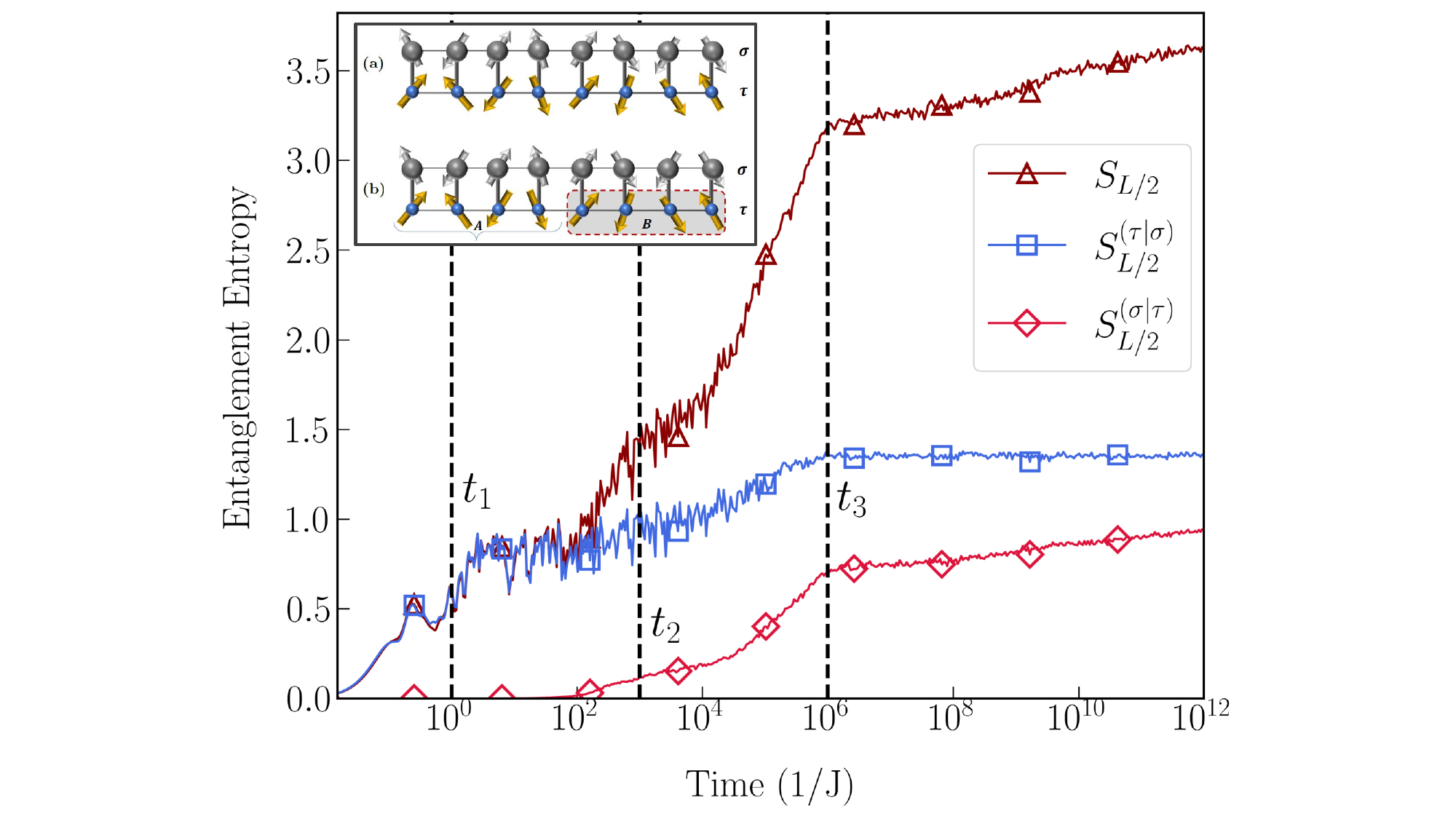}
	\caption{%
		(Color online)  \textbf{(Inset)} Schematic representation of the procedure used to evaluate the measured entanglement entropy.
        (a) A projective measurement is performed on the $\sigma$ spins (upper leg) of a many-body eigenstate, resulting in the collapse of the wavefunction to a configuration-dependent post-measurement state \( |\psi_p^{(m)}\rangle \) of the \(\tau\) spins.
        (b) For each measurement outcome, the half-cut entanglement entropy associated with subsystem \( A \) is computed as explained in the main text. \textbf{(Main)} Comparison of full and measured entanglement entropies in a spin ladder with strong inter-leg coupling \( J_z = 10 \).
		The total bipartite entropy \( S_{L/2} \) (dark-red triangles) displays three characteristic regimes as mentioned in Fig.~\ref{fig:ent_dyn}.
		Measured entropies after projective measurements on one leg, $S_A^{(\tau|\sigma)}$ (blue squares) and $S_A^{(\sigma|\tau)}$ (red diamonds), reveal the asymmetry in thermalization between legs.
		Notably, $S_A^{(\tau|\sigma)}$ saturates early, reflecting thermalization within the $\tau$, while  $S_A^{(\sigma|\tau)}$ remains strongly suppressed, indicating localization of the $\sigma$s.
	}
	\label{fig:measured_entropy_time}
\end{figure}

Upon performing a projective measurement on the \(\sigma\) spins in the computational (i.e., \( z \))-basis, the probability of obtaining the \(j\)-th configuration ($|\eta_j^{(\sigma)}\rangle$) in $\sigma$-Hilbert space is
\begin{equation}
	P_j^{(m)} = \sum_i |C_{ij}^{(m)}|^2.
\end{equation}
A schematic illustration is given in the inset-(a) of Fig.~\ref{fig:measured_entropy_time}, where the $\sigma$ configuration is fixed after the projection and is shown with grey color, while the linear combination
of the $\tau$ degrees of freedom define the projected state.
The resulting projected state on the \(\tau\) spins becomes
\begin{equation}
	|\psi_j^{(m)}\rangle = \frac{1}{\sqrt{P_j^{(m)}}} \sum_i C_{ij}^{(m)}\, |\mu_i^{(\tau)}\rangle,
\end{equation}
where the subscript $j$ stands for the  \(j\)-th configuration of $\sigma$ spins.
We then partition the \(\tau\) spins into two equal spatial regions \( A \) and \( B \), and compute the reduced density matrix of region \( A \) by tracing out \( B \):
\begin{equation}
	\rho_{A}^{(m,j)} = \tr_B\left[ |\psi_j^{(m)}\rangle \langle \psi_j^{(m)}| \right].
\end{equation}
An illustration of the bipartition of the $\tau$ spins is shwon in the inset-(b) of Fig.~\ref{fig:measured_entropy_time}, where the shadow on part B means the trace operation.
The entanglement entropy of region \( A \) for the \(j\)-th configuration is given by
\begin{equation}
	S_{A}^{(m,j)} = -\tr\left[ \rho_{A}^{(m,j)} \ln \rho_{A}^{(m,j)} \right].
\end{equation}

Finally, the full measured entanglement entropy of eigenstate \( |\psi^{(m)}\rangle \) is obtained by averaging over all measurement outcomes on the \(\sigma\) spins:
\begin{equation}
	S_{A}^{(\tau|\sigma)}(m) = \sum_j P_j^{(m)}\, S_{A}^{(m,j)}.
\end{equation}

This diagnostic captures the residual entanglement within the \(\tau\) spins after performing projective measurements on the \(\sigma\) spins. In thermal (ETH) eigenstates, \( S_{A}^{(\tau|\sigma)} \) remains volume-law, whereas in QDL regimes it drops significantly, indicating localization of lighter species and suppressed entanglement transfer across the legs. That means \( S_{A}^{(\tau|\sigma)} \) shows an area-law or sub-volume law in the QDL phase, indicating the localizatin of fast $\tau$ degrees of freedom
on the slow $\sigma$ partners.

The procedure is repeated by exchanging the $\sigma \leftrightarrow \tau$ spins in the above approach. That means the measurement on $\tau$ degress of freedom and obtaing the bipartite entanglement of the $\sigma$ chain, which results in \( S_{A}^{(\sigma|\tau)} \).

We present in Fig.~\ref{fig:measured_entropy_time} the time evolution of the entanglement entropy $S_{L/2}$ for a ladder system with $L=8$ rungs at strong coupling ($J_z = 10$). For times $t > t_3$, $S_{L/2}$ exhibits a clear logarithmic growth, indicative of quasi-MBL behavior(the same as shown in Fig.~\ref{fig:ent_dyn}).

In addition to $S_{L/2}$, we compute the measured entropies as previously defined. The entropy of $\tau$ spins conditioned on measured configurations of $\sigma$ spins, denoted by $S_A^{(\tau|\sigma)}$, is shown in blue with square markers. This entropy follows a similar temporal profile to $S_{L/2}$ up to $t_2$, and for $t>t_2$ displaying logarithmic growth and rapid saturation at $t_3$. This behavior suggests that the light species (i.e., $\tau$ spins) thermalizes quickly and does not retain memory of observables, contrary to the expectations of the QDL paradigm (see \hyperref[app:scaling]{Appendix~\ref*{app:scaling}}).

\begin{figure*}[t]
	\centering
	\begin{overpic}[width=0.3\linewidth]{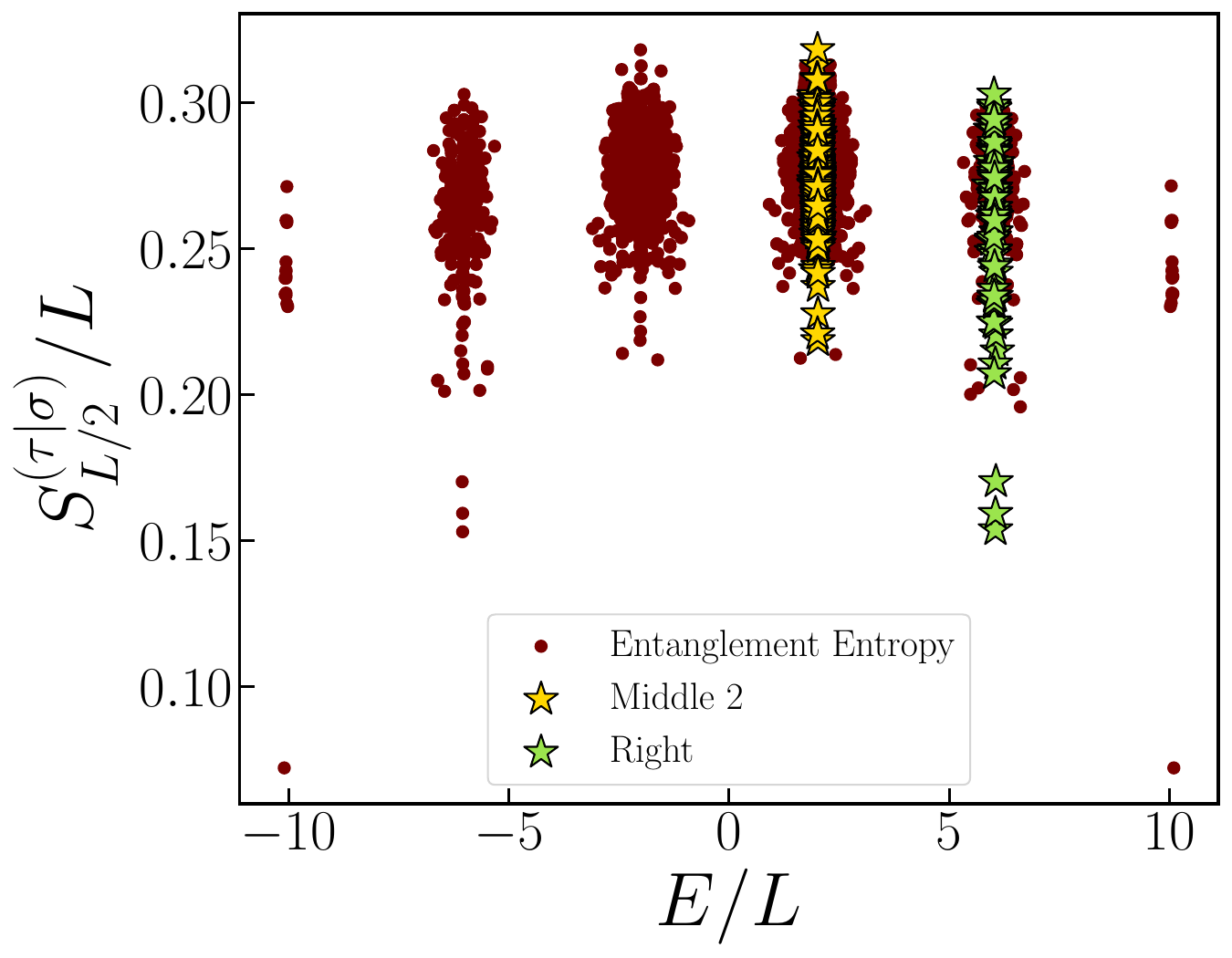}
		\put(21,66){\bfseries (a)} 
	\end{overpic}
	\hspace{2mm}
	\begin{overpic}[width=0.32\linewidth]{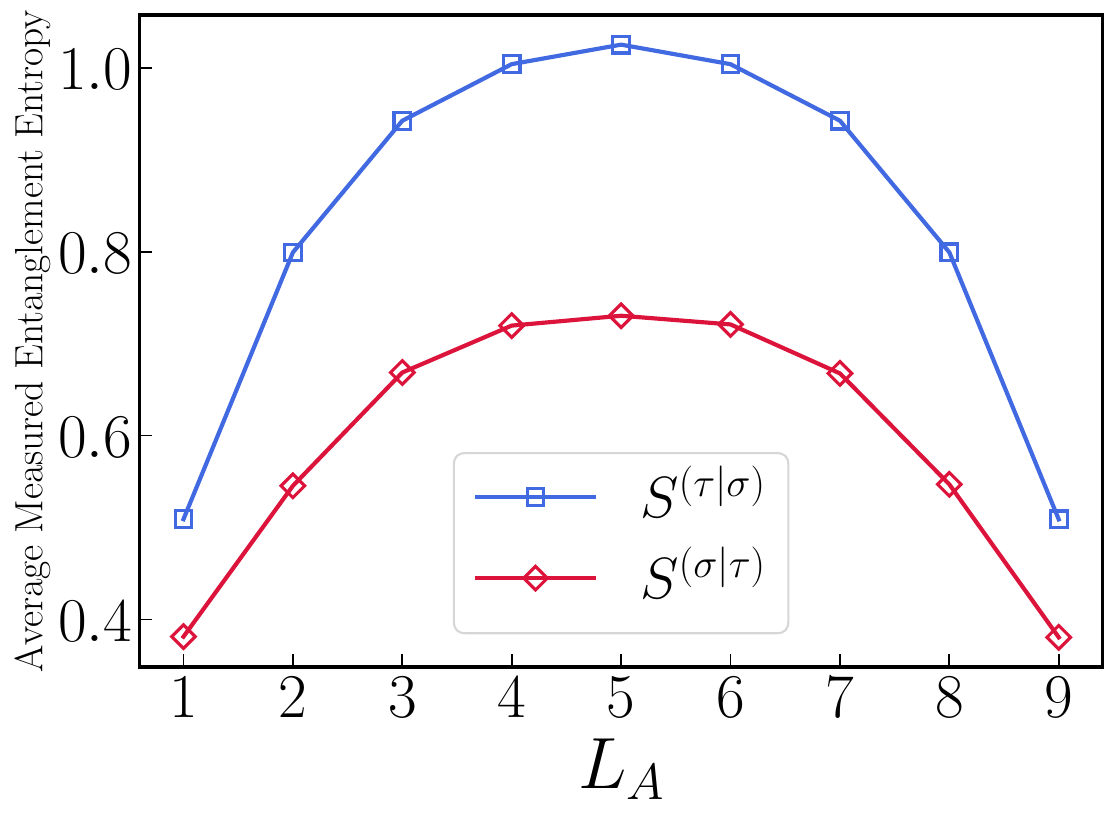}
		\put(14,64){\bfseries (b)} \put(32,50){\textcolor{black}{Middle-right band}}
	\end{overpic}
	\hspace{2mm}
	\begin{overpic}[width=0.32\linewidth]{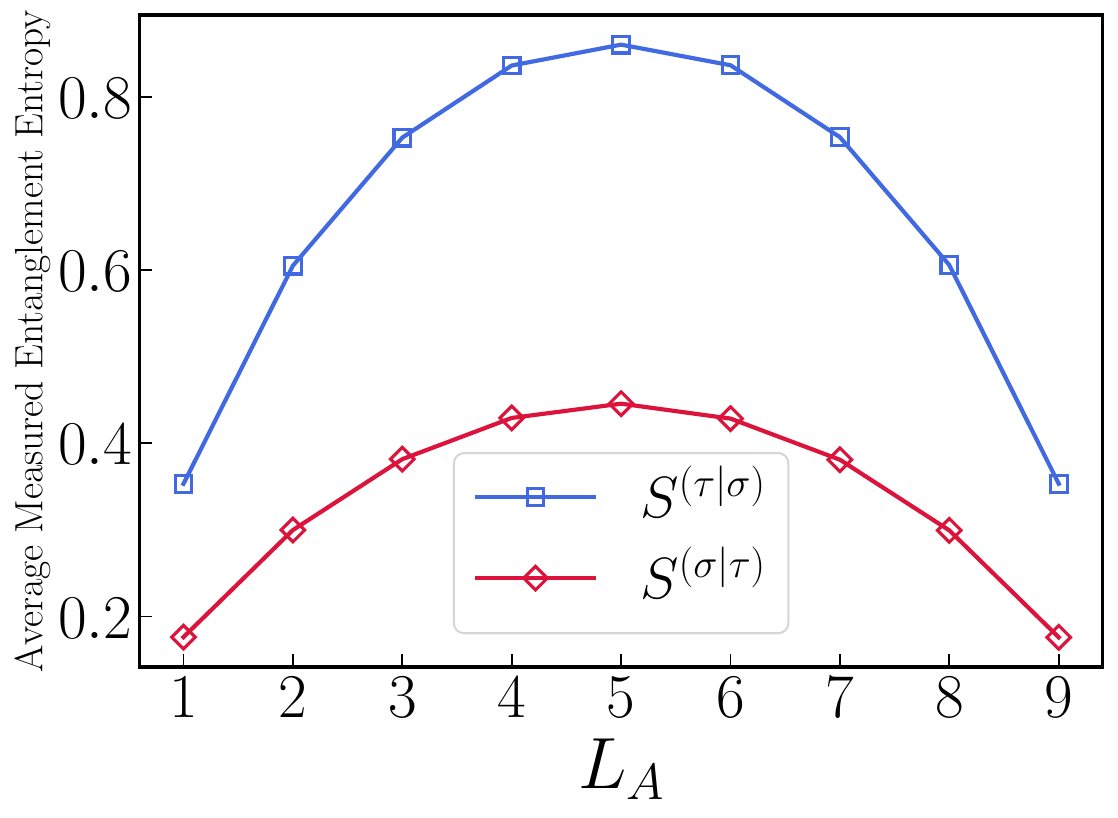}
		\put(14,64){\bfseries (c)}\put(40,50){\textcolor{black}{Right band}}
	\end{overpic}
	
	\caption{(Color online) (a) Energy-resolved measured entanglement entropy normalized by system size,
		\(S^{(\tau|\sigma)}_{L/2}/L\), versus energy density \(E/L\)
		 for a ladder of length $L=10$. 
		(b, c) Scaling of measured entropies $S_{L_A}^{(\tau|\sigma)}$ (blue, upper) and $S_{L_A}^{(\sigma|\tau)}$ (red, lower) with subsystem size $L_A$ for states in the middle-right and right bands, respectively.
		The $\tau$ spins exhibit near volume-law behavior, indicating rapid thermalization, while $\sigma$ spins show sub-volume scaling as a quasi-MBL phase.
}
	\label{fig:symmetric_bands}
\end{figure*}

Conversely, the entropy of $\sigma$ spins conditioned on measured $\tau$ configurations, $S_A^{(\sigma|\tau)}$, remains nearly frozen for $t < t_2$, reflecting extremely slow dynamics in this regime. Between $t_2$ and $t_3$, $S_A^{(\sigma|\tau)}$ increases rapidly, followed by an unbounded slow logarithmic growth for $t > t_3$. This evolution is responsible for the quasi-MBL behavior observed in $S_{L/2}$.

To further examine the contributions of $\tau$ and $\sigma$ spins to the entanglement dynamics of the model, we analyze the scaling behavior of the measured entropies of eigenstates with respect to subsystem size. As a first step, we plot the entanglement spectrum $S_{L/2}^{(\tau|\sigma)}$ as a function of energy per rung, $E/L$, in Fig.~\ref{fig:symmetric_bands}(a).

Limited by available computational resources,
we work at fixed magnetizations
using PBC and exploit discrete symmetries of the ladder—lattice translation \(\hat T\) and leg parity \(\hat{\mathcal P}\), i.e. $(\mathcal{S}^z_{\tau},\mathcal{S}^z_{\sigma},k,\mathcal{P})=(0, 0, 0, +1)$. The calculations are performed on a ladder system of length $L = 10$.
Projecting onto these sectors block–diagonalizes the Hamiltonian and greatly reduces the effective Hilbert spaces.
Within each block we compute, for all eigenstates available, the measured entropy of the light leg,
\(S^{(\tau|\sigma)}_{L/2}\), and display it against the energy density \(E/L\) in
Fig.~\ref{fig:symmetric_bands}-(a)(An analogous plot for the heavy leg, \(S^{(\sigma|\tau)}_{L/2}\), displays the same band segmentation and is omitted).  Four dominant bands are clearly resolved; two narrow edge bands exist but are not analyzed further.

For the scaling analysis in panels (b,c) we \emph{do not} impose symmetry resolution in order to observe the generic behavior: instead, we target narrow spectral windows centered in the middle–right and right bands (starred regions in panel (a)) using a shift-and–invert method on the full (imposing \(\mathcal{S}^z_{\tau}=\mathcal{S}^z_{\sigma}=0\)) Hilbert space.  From each band, we sample the central \(20\%\) of eigenstates (by energy), evaluate the measured entropies \(S^{(\tau|\sigma)}_{L_A}\)(blue, upper curves) and \(S^{(\sigma|\tau)}_{L_A}\)(red, lower curves) for subsystems of length \(L_A\), and report the band averages as functions of \(L_A\) in Figs.~\ref{fig:symmetric_bands}(b)–(c). In both panels, $S_{L_A}^{(\sigma|\tau)}$ exhibits sub-volume scaling, $S_{L_A}^{(\tau|\sigma)}$ displays significantly larger values and a stronger dependence on $L_A$, indicative of volume-law scaling. The left bands in Fig.~\ref{fig:symmetric_bands}-(a) are related to the right bands by symmetry and therefore exhibit identical behavior; they are omitted for brevity.

These observations are consistent with previous diagnostics: the $\sigma$ spins exhibit slow thermalization, retaining memory and gradually saturating their entropy in a logarithmic-in-time fashion. Meanwhile, the $\tau$ spins thermalize rapidly, as evidenced by their volume-law scaling of entanglement entropy.

We therefore conclude that the observed slow logarithmic growth of entanglement at $t>t_3$ cannot be attributed to the standard QDL mechanism. Instead, it resembles a reversed-QDL phase. In the following section, we argue that this behavior originates from the emergence of local integrals of motion in the strong coupling limit ($J_z \gg 1$) of the model.

\subsection{Emergent XXZ Dynamics in the strong coupling limit}
\label{sec:effective hamiltonian}
To capture the rich physics of the strongly coupled ladder in the regime \( J_z \gg J, J' \), we derive effective Hamiltonians for both the \textbf{low-energy} and \textbf{high-energy} subspaces.
These correspond to two mutually orthogonal doublets localized on each rung of the ladder, with each doublet comprising a pair of degenerate eigenstates of the corresponding rung Hamiltonian.

\noindent\textit{Low-energy effective theory}—We begin by projecting the full Hamiltonian onto the lower-energy doublet states of all rungs \( \{ \ket{\gamma_l^+}, \ket{\gamma_l^-}; l=1, \cdots L \} \), which define an emergent spin-\(\tfrac{1}{2}\) degree of freedom per rung. The corresponding Pauli operators \( \Gamma_l^\alpha \) (\( \alpha = x, y, z \)) obey the standard algebra and act within this subspace.\\
The first-order contribution vanishes due to orthogonality with the excited sector, so we proceed with the second-order Brillouin–Wigner perturbation theory~\cite{BLOCH1958329, Winkler2003,Bravyi2011,MacDonald1988,CohenTannoudji1992,Hubac2000}. The resulting effective Hamiltonian is (for details see \hyperref[app:eff_Hamiltonian]{Appendix~\ref*{app:eff_Hamiltonian}}):
\begin{align}
	\mathcal{H}_{\text{eff}}^{\text{(low)}} &=\
	 -\frac{2 J J'}{J_z}
	\sum_{l=1}^L \left(
	\Gamma_\ell^x \Gamma_{\ell+1}^x + \Gamma_\ell^y \Gamma_{\ell+1}^y
	\right) \notag \\
	 & + (\frac{J^2 + J'^2}{J_z}) \sum_{l=1}^L \Gamma_\ell^z \Gamma_{\ell+1}^z
	- \left( \frac{J^2 + J'^2}{J_z} \right) \mathbb{1}.
\end{align}
This Hamiltonian describes an effective XXZ spin chain, where transverse couplings arise from flip-flop processes and the longitudinal interaction originates from virtual excursions into the excited subspace. The final constant term induces a uniform shift in energy. The anisotropy parameter of the Heisenberg spin chain, defined as the ratio between the longitudinal (\(zz\)) and transverse (\(xy\)) coupling terms, satisfies the condition
\begin{equation}
\Delta = \frac{J^2 + J'^2}{2JJ'} \geq 1.
 \label{Delta}
\end{equation}
This inequality places the system in the Ising regime, where the model remains integrable.

\vspace{1em}
\noindent\textit{High-energy effective theory}—The high-energy subspace is spanned by the excited doublets of rungs \( \{ \ket{\varphi_l^+}, \ket{\varphi_l^-}; l=1, \cdots L  \} \). Defining a new set of pseudospin operators \( \hat{\Phi}_l^\alpha \) within this subspace:
\begin{align}
	\hat{\Phi}_l^+ &= \ket{\varphi_l^+} \bra{\varphi_l^-}, \notag\\
	\hat{\Phi}_l^- &= \ket{\varphi_l^-} \bra{\varphi_l^+}, \notag \\
	\hat{\Phi}_l^z &= \ket{\varphi_l^+} \bra{\varphi_l^+} - \ket{\varphi_l^-} \bra{\varphi_l^-}.
\end{align}
We again apply the second-order Brillouin–Wigner perturbation theory, yielding the effective Hamiltonian:
\begin{align}
	\mathcal{H}_{\text{eff}}^{\text{(high)}} &=\
	 \frac{2 J J'}{J_z}
	\sum_{l=1}^L \left(
	\hat{\Phi}_\ell^x \hat{\Phi}_{\ell+1}^x + \hat{\Phi}_\ell^y \hat{\Phi}_{\ell+1}^y
	\right) \notag \\
	& - (\frac{J^2 + J'^2}{J_z})
	\sum_{l=1}^L \hat{\Phi}_\ell^z \hat{\Phi}_{\ell+1}^z
	+ \left(\frac{J^2 + J'^2}{J_z} \right) \mathbb{1}.
\end{align}
This sector also realizes an XXZ-type chain, but with \textbf{opposite sign} in the longitudinal interaction, reflecting its distinct spectral origin. The positive constant term shifts the high-energy manifold upward. Together, the two effective theories offer a complete picture of emergent spin dynamics across energy scales in the ladder system. The anisotropy of the high-energy effective Heisenberg chain satisfies Eq.~\ref{Delta}, which puts the system into Ising regime.

The energy separation between the two doublets is given by \(2J_z\), rendering them energetically well-separated in the strong coupling regime. In this limit, the effective theory of the spin ladder indicates that the Hilbert space decomposes into two dynamically disconnected sectors. Each sector is effectively described by an integrable spin-\(\frac{1}{2}\) Heisenberg chain with anisotropy \(\Delta > 1\), placing the system within the Ising regime. Consequently, for \(J_z \gg J, J'\), the full model admits a description in terms of emergent constants of motion associated with the integrable Heisenberg chains. This structure provides a natural interpretation of the quasi-MBL behavior observed at large \(J_z\), where the local integrals of motion originate from the underlying integrability of the effective chains.

\section{Summary and conclusions}
\label{Summary}

\begin{figure}[t]
	\centering
	\includegraphics[width=1.\linewidth]{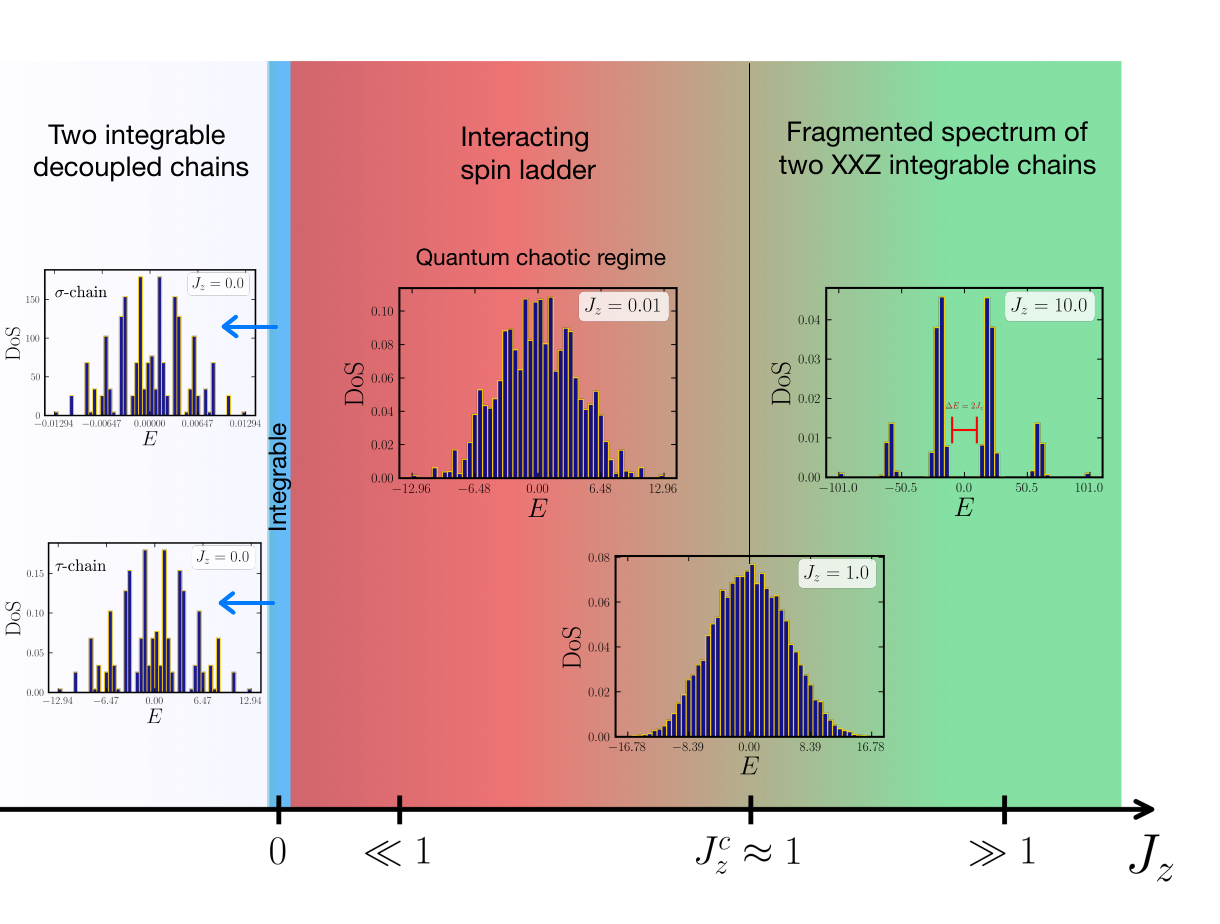}  
	\caption{%
		(Color online) Phase diagram of the spin ladder as a function of the inter-leg Ising coupling \( J_z \).
		For \( J_z = 0 \), the system consists of two decoupled integrable chains (blue region), each exhibiting a discrete and regular spectrum. 
		As \( J_z \) increases (\( J_z \ll 1 \)), the inter-leg interactions induce chaotic dynamics, leading to a quantum chaotic regime (red region). 
		Notably, the density of states for small \( J_z \) (e.g., \( J_z = 0.01 \)) actually signals the onset of coupling between the two chains, giving rise to quantum ergodicity and ETH-like behavior.
		Around the crossover scale \( J_z^c \sim 1 \), the density of states (DoS) becomes fully broadened and ergodic. 
		For \( J_z \gg 1 \), the ladder effectively reduces to two weakly coupled XXZ chains in the strong coupling limit (green region), exhibiting a fragmented spectrum arising from the emergent effective theory derived in Sec.~\ref{sec:effective hamiltonian}. 
		The spectrum consists of a series of subbands, each separated by an energy gap of \( \Delta E = 2J_z \gg 1 \).
		Inset plots show representative DoS for selected values of \( J_z \).
	}
	\label{fig:phase_diagram}
\end{figure}

In this work, we have investigated the long-time dynamics of the two-leg spin-$\tfrac{1}{2}$ ladder introduced in~\cite{Yao2016}, consisting of asymmetric XY interactions along the legs ($J \gg J'$) and an Ising interaction on the rungs (coupling $J_z$). Our results demonstrate that the dynamical behavior of the system is highly sensitive to $J_z$, giving rise to the phase diagram presented in Fig.~\ref{fig:phase_diagram}. At $J_z=0$, the system decouples into two independent spin-$\tfrac{1}{2}$ XY chains, which are integrable and exhibit the characteristic features of integrable models. The density of states (DoS) in this limit consists of two identical contributions with distinct energy scales reflecting the leg asymmetry ($J=1$, $J'=0.001$). Introducing a finite rung coupling $J_z$ induces interactions between the chains and drives the system into a chaotic regime.
This transition manifests itself through the broadening of the density of states (DoS) and the emergence of statistical features consistent with the eigenstate thermalization hypothesis (ETH).
This regime is characterized by rapid growth of the half-cut entanglement entropy toward its saturation value following a volume-law scaling, exponential growth of the rescaled adiabatic gauge potential norm with system size, and level statistics approaching the Gaussian orthogonal ensemble (GOE) limit. The corresponding DoS evolves into a compact distribution, as illustrated in Fig.~\ref{fig:phase_diagram} for $J_z=0.01$. The chaotic phase persists up to a critical coupling $J_z^c$.

At the critical rung coupling $J_z=J_z^c$, the system undergoes a dynamical crossover to a reversed-QDL regime. Numerical results place the value of $J_z^c \approx 1.0$, as inferred from several diagnostics: a pronounced peak in fidelity susceptibility, a crossover in level statistics from GOE to Poisson, a reduction of eigenstate entropies, and a change in the scaling of the rescaled AGP norm with system size. The DoS at the crossover displays a compact profile, consistent with critical behavior. Due to strong finite-size effects and the exponential growth of Hilbert space dimension with ladder length, we were limited to ladders of length $L=12$ ($N=24$ spins). For smaller sizes ($L<10$), level statistics remain below the Poisson limit for almost all $J_z$, highlighting the need for larger system sizes to establish scaling behavior with higher precision.

For strong couplings ($J_z \gg 1$), the system exhibits a reversed-QDL regime, as evidenced by logarithmic growth of the entanglement entropy, super-linear scaling of the rescaled AGP norm with system size, and Poissonian level statistics. To explore the underlying mechanism, we examined the QDL scenario. In the highly asymmetric limit ($J' \ll J$), the upper and lower legs may be regarded as slow and fast fermionic modes, respectively. While the standard QDL picture suggests thermalization of the slow modes and localization of the fast ones, our analysis reveals the opposite: the upper spins (slow fermions) remain localized, while the lower spins thermalize rapidly. This reversed-QDL behavior is reflected in the measurement-based entanglement entropy, which shows logarithmic growth of the upper-spin entropy when conditioned on lower-spin measurements (Fig.~\ref{fig:measured_entropy_time}). Furthermore, a second-order effective Hamiltonian in the strong-coupling regime reduces to two decoupled spin-$\tfrac{1}{2}$ Heisenberg chains with spectra separated by $\Delta E = 2J_z \gg 1$. This spectral fragmentation, evident in the DoS at $J_z=10$ (Fig.~\ref{fig:phase_diagram}), provides a natural source of local integrals of motion stabilizing the quasi-MBL phase.

Finally, we emphasize that the asymmetry of the leg couplings plays a crucial role. At $J_z=10$, increasing $J'$ toward $J$ leads to rapid growth of the entanglement entropy and restores thermalization, even in the strong-coupling regime. This confirms that the imbalance between the legs is essential for sustaining the quasi-MBL behavior. Taken together, these results establish a rich dynamical phase structure for the two-leg spin-$\tfrac{1}{2}$ ladder: an integrable regime at $J_z=0$, a chaotic thermal phase for $0<J_z<J_z^c$, and a quasi-MBL regime for $J_z\gtrsim J_z^c$, with the strong-coupling limit exhibiting reversed-QDL behavior and spectral fragmentation.

The ladder Hamiltonian exhibits extensive symmetries—most notably conservation of magnetization on each leg, translation invariance, and spin-inversion symmetry—which necessitate working in carefully resolved subspaces. 
Additionally, the presence of two markedly distinct energy scales in our model induces significant finite-size effects, making it challenging to extrapolate numerical results from small systems to the thermodynamic limit ~\cite{Papi2015}.
Moreover, the exponential growth of the Hilbert space dimension restricts us to system sizes up to $L=12$, which, although sufficient to expose clear dynamical trends, leaves open the question of scaling to the thermodynamic limit.

In summary, we demonstrate a reentrant transition with increasing $J_z$: from an integrable nonergodic regime, through a chaotic thermalizing phase, and into a robust nonthermal regime at strong coupling. The latter hosts a reversed QDL structure, where the light species thermalizes while the heavy one remains localized, establishing a disorder-free mechanism for localization driven by inter-sector entanglement asymmetry. Our results thus provide a minimal microscopic route to interaction-induced nonergodicity, with implications for constrained dynamics, quantum simulators, and the broader taxonomy of nonthermal quantum matter.

\begin{acknowledgments}
	This work was supported by the Iran National Science Foundation (INSF) under Grant No.~4037050. 
	The authors thank Hyeongjin Kim for sharing helpful material. Special thank is devoted to H. Yarloo for the careful reading of the manuscript, fruitful discussions and useful comments.
	Part of the numerical computations was performed using the QuSpin library \cite{QuSpin2017,QuSpin2019}.
\end{acknowledgments}

\appendix

\section{Jordan-Wigner transformation}
\label{Jordan-Wigner}

Although we do not numerically simulate the fermionic model of Eq.~\eqref{Fermion_H}, we briefly explain the Jordan-Wigner transformation of the original spin model (Eq.~\eqref{eq:H}) to its fermionic counterpart. According to the Jordan-Wigner transformation~\cite{Jordan1928,Lieb1961}, the Pauli operators on the lower leg ($\tau$-spins) are defined by
\begin{eqnarray}
\label{JWT}
\tau^+_{\ell}&=&c_{\ell}^{\dagger} \; exp( i \pi \sum_{j< \ell} n_j ), \nonumber \\
\tau^-_{\ell}&=& exp( -i \pi \sum_{j<\ell} n_j ) \; c_{\ell} , \\
\tau^z_{\ell}&=&2(n_{\ell} - \frac{1}{2}) \;\;, \;\; n_{\ell}=c_{\ell}^{\dagger}c_{\ell}. \nonumber
\end{eqnarray}
Accordingly, we get
\begin{equation}
\tau^+_{\ell} \tau^-_{\ell+1} + \tau^-_{\ell} \tau^+_{\ell+1} = 
c_{\ell}^{\dagger} c_{\ell+1} + c_{\ell+1}^{\dagger} c_{\ell}, 
\end{equation}
for $\ell=1, \cdots, L-1$, except the boundary terms.
The corresponding boundary terms for the lower leg are given by:
\begin{equation}
\label{btau}
\tau^+_{L} \tau^-_{1} + \tau^-_{L} \tau^+_{1} = -F(c_{L}^{\dagger} c_1 + c_1^{\dagger} c_L),
\end{equation}
where $F$ is the Klein factor
\begin{equation}
F=exp(i \pi \hat{N}) \;\;\;; \;\;\; \hat{N}=\sum_{\ell =1}^{L} n_{\ell}.
\end{equation}
A similar transformation for the upper leg (\(\sigma\)-spins) is given by the following equations, where the inclusion of the Klein factor is responsible for preserving the commutation relations of the original spin operators,
\begin{eqnarray}
\label{JWS}
\sigma^+_{\ell}&=&\tilde{c}_{\ell}^{\dagger} \; exp( i \pi \sum_{j< \ell} \tilde{n}_j ) \; F, \nonumber \\
\sigma^-_{\ell}&=& F^{\dagger} \; exp( -i \pi \sum_{j<\ell} \tilde{n}_j )\; \tilde{c}_{\ell} , \\
\sigma^z_{\ell}&=&2(\tilde{n}_{\ell} - \frac{1}{2}) \;\;, \;\; \tilde{n}_{\ell}=\tilde{c}_{\ell}^{\dagger}\tilde{c}_{\ell}. \nonumber
\end{eqnarray}
After some calculations, we arrive at the following boundary term for the upper leg:
\begin{equation}
\label{bsigma}
\sigma^+_{L} \sigma^-_{1} + \sigma^-_{L} \sigma^+_{1} = - \tilde{F} (\tilde{c}_{L}^{\dagger} \tilde{c}_1 +  \tilde{c}_1^{\dagger} \tilde{c}_L),
\end{equation}
where,
\begin{equation}
\tilde{F}=exp(i \pi \hat{\tilde{N}}) \;\;\;; \;\;\; \hat{\tilde{N}}=\sum_{\ell =1}^{L} \tilde{n}_{\ell}.
\end{equation}
Both $F$ and $\tilde{F}$ yield $\pm 1$ depending on the parity of fermion occupations of the lower and upper legs, respectively.
The boundary terms in Eqs.\eqref{btau} and \eqref{bsigma} demonstrate that an odd (even) parity of fermion occupations in either the legs of ladder necessitates a periodic (anti-periodic) boundary term for the fermionic model. Finally, we arrive at the fermionic Hamiltonian, Eq.~\eqref{Fermion_H}.

\section{Level spacing statistics}
\label{app:level-spacing}

We compute the mean level-spacing ratio to distinguish between integrable
(non-thermal) and chaotic (thermal) regimes~\cite{Oganesyan2007,Atas2013}.
All numerical exact diagonalizations are performed in the symmetry-resolved sector
$(\mathcal{S}^z_{\tau}=0, \mathcal{S}^z_{\sigma}=0, k, \mathcal{Z})$.
For $L=10$ all sixteen symmetry sectors are considered, while for
$L=12$ six symmetry sectors $(0,0,\pi/6,\pm 1)$, $(0,0,5\pi/6,\pm 1)$ and $(0,0,11\pi/6,\pm 1)$ are
taken into account. To avoid the edge effects we restrict the computations to the
50\% of the levels in the centre of spectrum for both $L=10$ and $L=12$.

The mean ratio is defined as
\begin{equation}
	r_n = \frac{\min(\delta_n, \delta_{n+1})}{\max(\delta_n, \delta_{n+1})},
\end{equation}
where $\delta_n = E_{n+1} - E_n$ denotes the spacing between consecutive
energy levels. A key advantage of the $r$-statistic is that it does not
require spectral unfolding, which makes it particularly robust in finite-size
systems. The ensemble average $\langle r \rangle$ approaches
$\approx 0.531$ for Wigner--Dyson (chaotic) statistics and
$\approx 0.387$ for Poisson (integrable) statistics~\cite{Atas2013}.

\begin{figure}[t]
	\centering
	\includegraphics[width=1.0\linewidth]{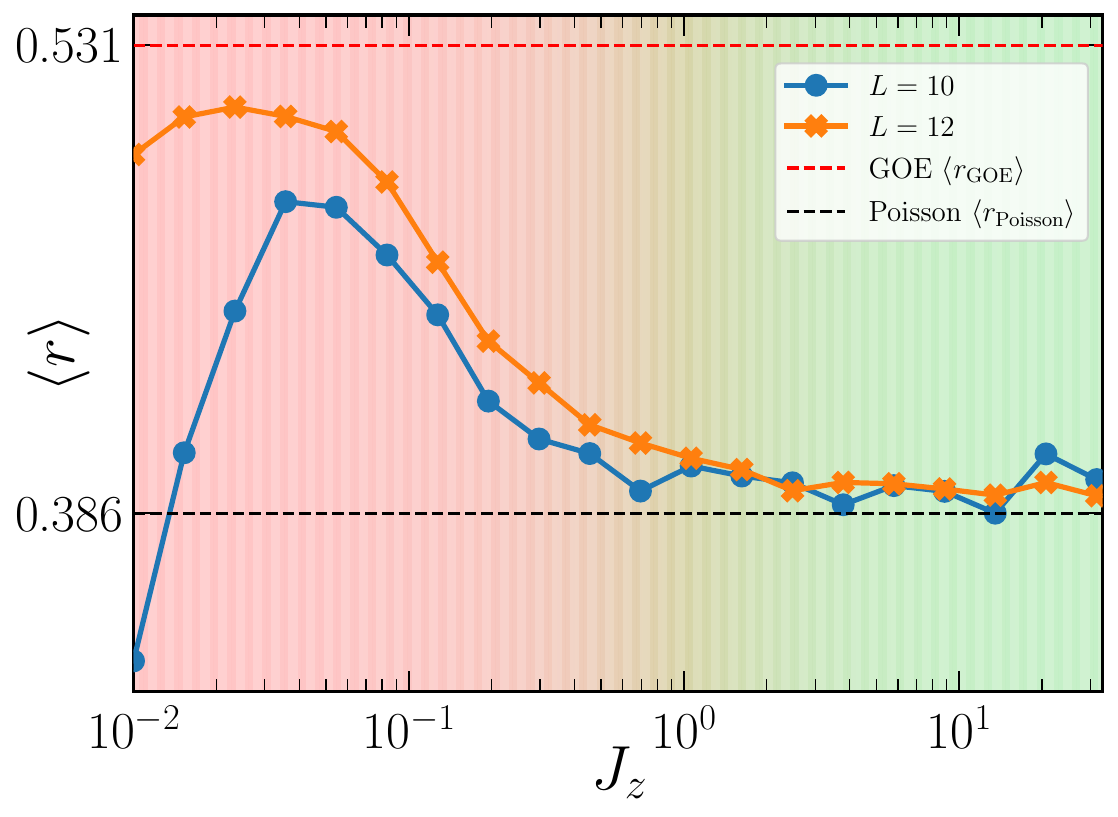}
		
		\caption{(Color online)  Average adjacent level-spacing ratio $\langle r \rangle$ as a function of $J_z$ for system sizes $L=10$ and $L=12$.
			Increasing $J_z$ induces a crossover around $J_z\simeq 1$ from GOE (choatic) to Poisson statistics. 
			See the main text for the details.
		}
		\label{fig:mean_gap_ratios_all_L}
	\end{figure}

Figure~\ref{fig:mean_gap_ratios_all_L} shows the average ratio $\langle r\rangle$ as a function of $J_z$ for system sizes $L=10$ and $L=12$.
At $J_z=0$, the model reduces to two decoupled integrable spin-$\tfrac{1}{2}$ XY chains, yielding the Poissonian value of $\langle r\rangle$.
Increasing $J_z$ induces a dynamical transition into a chaotic phase, where $\langle r\rangle$ approaches the Gaussian Orthogonal Ensemble (GOE) value.
For both $L=10$ and $L=12$, the data clearly exhibit a sharp transition from the Poissonian limit
to the GOE regime as $J_z$ is increased.

As $J_z$ increases further, the mean level-spacing ratio $\langle r\rangle$ decreases again toward the Poisson value around $J_z\simeq 1$.
This trend is consistent with other diagnostics indicating that a dynamical crossover back to Poisson-like statistics occurs at large $J_z$,
signaling MBL-like behavior in this regime.

It is important to note that the system under study is quasi-1D (a two-leg ladder). Consequently, extending calculations beyond $L=12$ (corresponding to more than 24 spins) requires computational resources beyond typical availability. Additionally, results for smaller system sizes ($L<10$) are not presented due to pronounced finite-size effects although confirm the qualitative behavior.

\section{Scaling fits as evidence for a \textit{reversed} QDL regime}
\label{app:scaling}

\begin{figure*}[t]
	\centering
	\setlength{\tabcolsep}{2pt}
	
	\begin{minipage}[t]{0.33\textwidth}
		\centering
		\begin{overpic}[width=\linewidth]
			{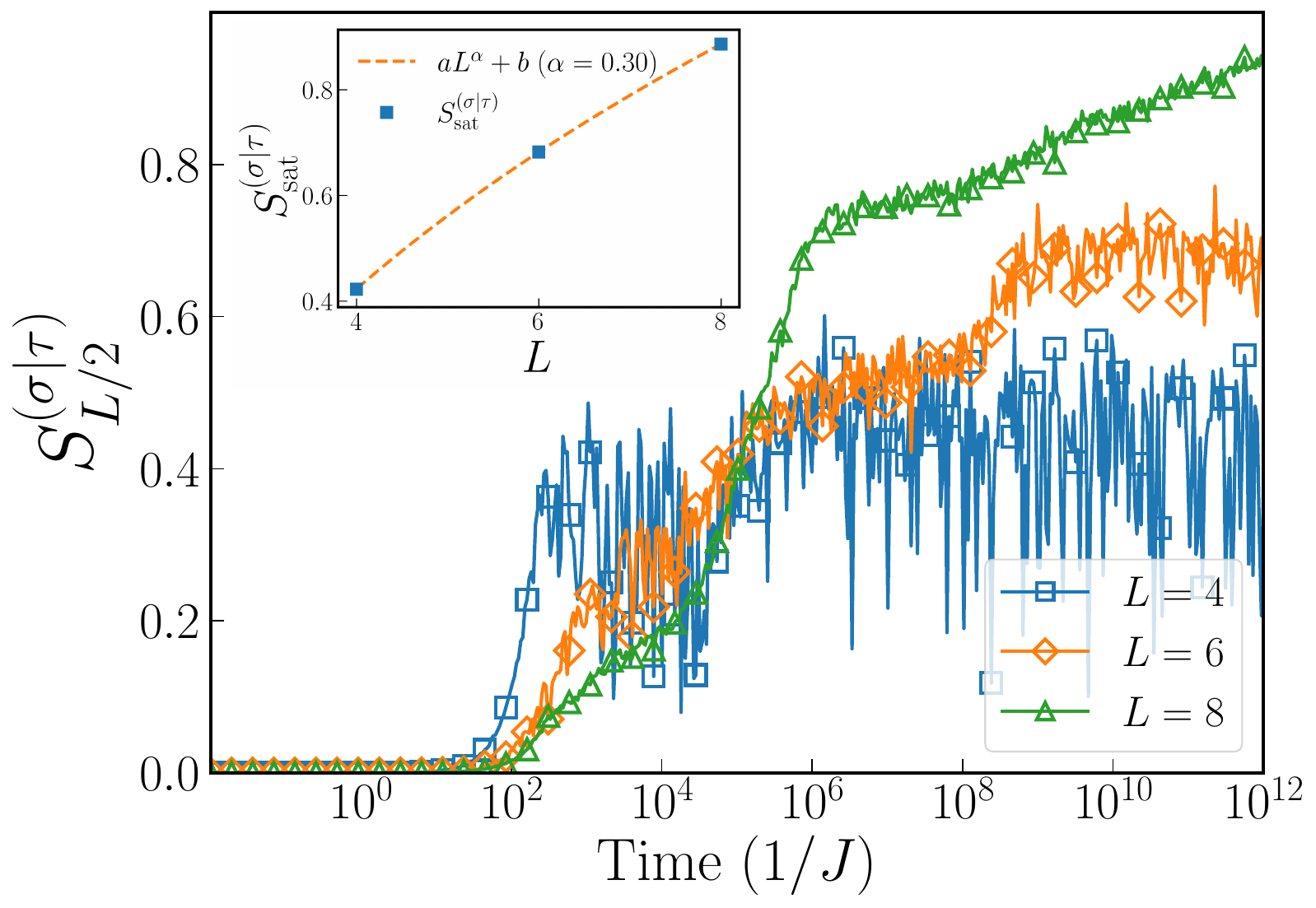}
			\put(18,40){\bfseries (a)}
		\end{overpic}
	\end{minipage}
	\begin{minipage}[t]{0.32\textwidth}
		\centering
		\begin{overpic}[width=\linewidth]
			{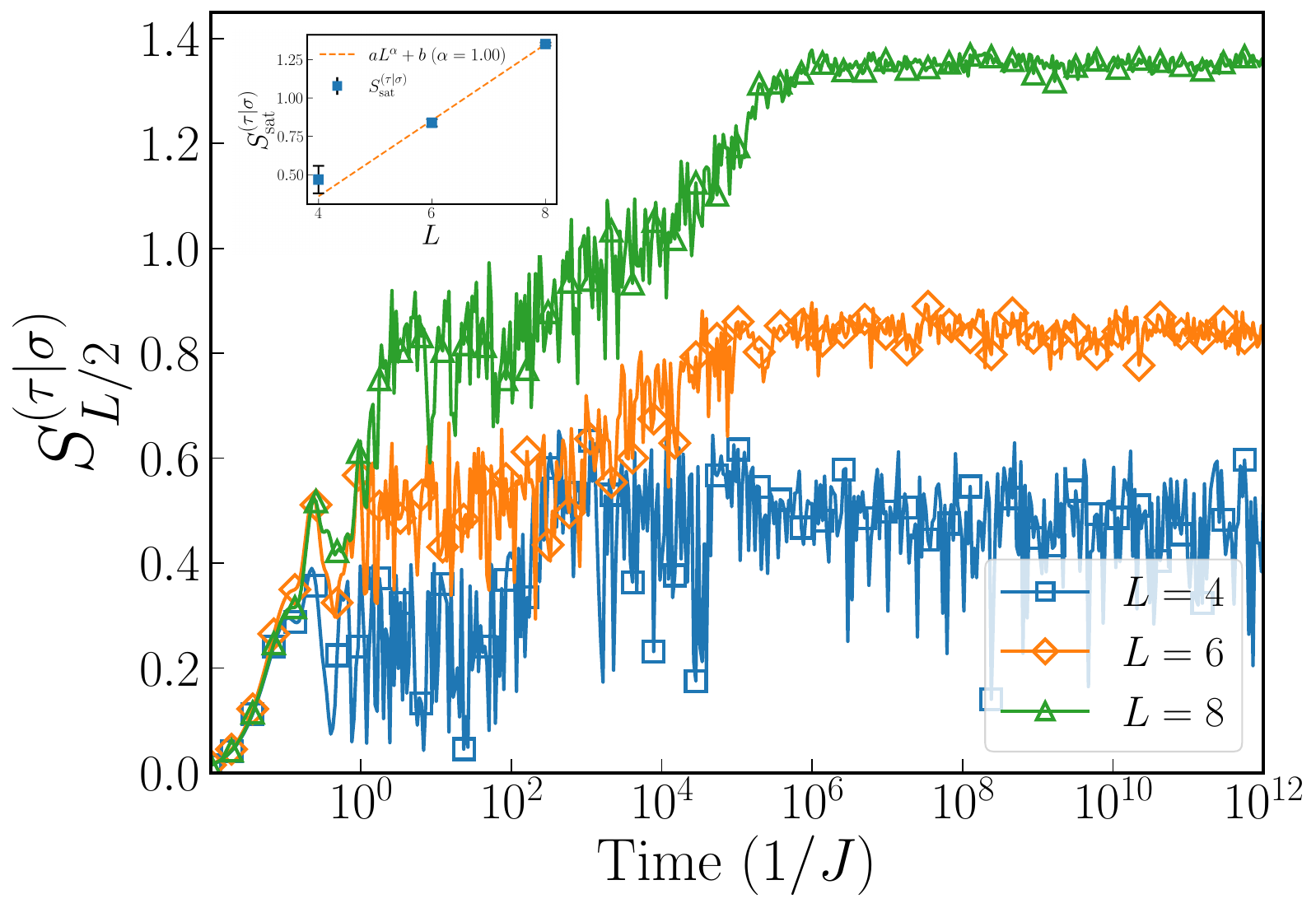}
			\put(18,40){\bfseries (b)}
		\end{overpic}
	\end{minipage}
	\begin{minipage}[t]{0.33\textwidth}
		\centering
		\begin{overpic}[width=\linewidth]
			{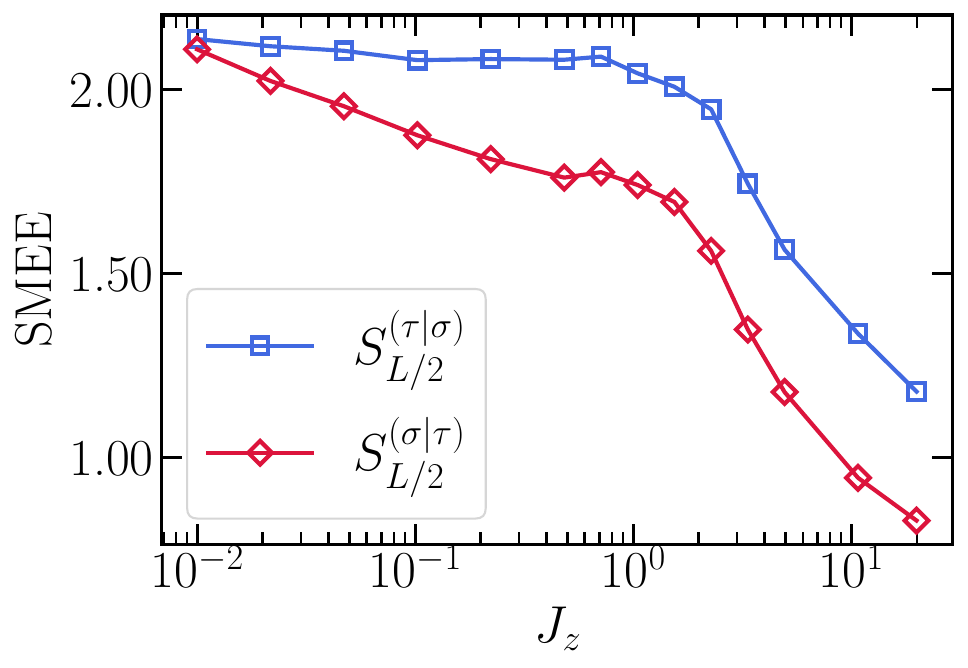}
			\put(20,42){\bfseries (c)}
		\end{overpic}
	\end{minipage}
	
	\vspace{-0.4cm}
	\caption{(Color online)
		(a) \textbf{Heavy leg (\(\sigma\))—measured half-cut entanglement entropy.}
		Main panel shows \(S^{(\sigma|\tau)}_{L/2}(t)\); the inset reports the terminal-window saturation values \(S^{(\sigma|\tau)}_{\rm sat}(L)\) together with the best \emph{power-law} fit \(S_{\rm sat}(L)=aL^{\alpha}+b\) with \(\alpha\simeq0.30\), indicating sub-volume growth. 
		(b) \textbf{Light leg (\(\tau\))—measured half-cut entanglement entropy.}
		Main panel shows \(S^{(\tau|\sigma)}_{L/2}(t)\); the inset displays \(S^{(\tau|\sigma)}_{\rm sat}(L)\) with the best linear fit \(S_{\rm sat}(L)=aL+b\), consistent with volume-law scaling. Error bars denote the one–standard-deviation spread of the terminal window used to define \(S_{\rm sat}\).
		(c) Long-time saturation values of the measured half-cut entanglement entropy versus \(J_z\) for \(L=8\), showing the separation between \(S^{(\tau|\sigma)}_{L/2}\) and \(S^{(\sigma|\tau)}_{L/2}\) at strong coupling.
		Results in (a) and (b) are shown for \(L=4,6,8\) with periodic boundary conditions. Model parameters are \(J=1\), \(J'=10^{-3}\), \(J_z=10\), and calculations are performed in the symmetry sector \(\mathcal{S}^z_{\sigma}=\mathcal{S}^z_{\tau}=0\).}
	
	\label{fig:timeseries_dual}
\end{figure*}

To diagnose scaling in our two–leg ladder, we analyze the long–time saturation values of the
\emph{measured} half-cut entanglement entropy \(S^{(\beta)}_{L/2}(t)\) at system sizes
\(L=4,6,8\), with \(\beta\in\{\tau|\sigma,\ \sigma|\tau\}\).

For each system size $L$ we record the time series $S^{(\beta)}_{L/2}(t)$ at
discrete sampling times $t_1,\dots,t_{M_L}$, where $M_L$ is the number of
time samples.

We estimate the long-time \emph{saturation values} by averaging over the last fraction
of this record:

\begin{itemize}
	\item Choose a tail fraction $f$ (we use $f=0.2$).
	\item Let $K_L=\lfloor f\,M_L\rfloor$ be the number of samples in the tail.
	\item Define the tail (terminal) index set
	$\mathcal{I}_{\rm tail}(L)=\{M_L-K_L+1,\dots,M_L\}$,
	i.e., the last $K_L$ points of the record.
\end{itemize}

The saturation value is then the mean of those last $K_L$ samples:
\begin{equation}
	S^{(\beta)}_{\rm sat}(L)
	= \frac{1}{K_L}\sum_{k\in\mathcal{I}_{\rm tail}(L)}
	S^{(\beta)}_{L/2}(t_k).
	\label{eq:Ssat-discrete}
\end{equation}

Its error bar reflects the scatter within that tail window. We report the
standard error of the mean (SEM),
\begin{equation}
	\mathrm{SEM}(L)=\frac{s(L)}{\sqrt{K_L}},\qquad
	s(L)^2=\frac{1}{K_L-1}\sum_{k\in\mathcal{I}_{\rm tail}(L)}
	\big[S^{(\beta)}_{L/2}(t_k)-S^{(\beta)}_{\rm sat}(L)\big]^2 .
\end{equation}

We then fit the saturation values by (weighted) least squares to three size–scaling ansätze:
\begin{align}
	\text{linear:}\quad & S_{\rm sat}(L)=a\,L+b, \label{eq:lin}\\[2pt]
	\text{power law:}\quad & S_{\rm sat}(L)= a\,L^{\alpha}+ b, \label{eq:power}\\[2pt]
	\text{logarithmic:}\quad & S_{\rm sat}(L)=c\,\ln L+d. \label{eq:log}
\end{align}
When SEM estimates are available, each datum is weighted by \(w_i=\mathrm{SEM}(L_i)^{-2}\).
We report the coefficient of determination \(R^2\) for goodness of fit, but model preference is
established using out–of–sample and information–theoretic criteria: residual sum of squares (RSS),
the small–sample–corrected Akaike information criterion AICc~\cite{Akaike1974,Hurvich1989} with
Akaike weights~\cite{Burnham2002}, leave–one–out cross–validated MSE (LOO–CV)~\cite{Stone1977},
and the Bayesian information criterion BIC~\cite{Schwarz1978}.
Lower AICc/BIC and CV–MSE, together with smaller RSS (and larger \(R^2\)), indicate the superior model.

\paragraph*{Heavy (\(\sigma\)) leg — sub–volume scaling.}
From the long–time saturation values [Fig.~\ref{fig:timeseries_dual}(a)], a two–parameter power law
\(S^{(\sigma|\tau)}_{\rm sat}(L)=aL^{\alpha}+b\) is decisively preferred by information–theoretic criteria over
logarithmic and linear alternatives. Using BIC we find
\(\mathrm{BIC}_{\rm pow}=-66.45\), \(\mathrm{BIC}_{\ln}=-25.47\), and
\(\mathrm{BIC}_{\rm lin}=-20.61\) (with comparable parameter counts),
while the fits have \(R^2_{\rm pow}=1.000\) vs.\ \(R^2_{\ln}=0.997\) and residual sums of
squares \(\mathrm{RSS}_{\rm pow}=3.47\times10^{-10}\) vs.\ \(\mathrm{RSS}_{\ln}=2.97\times10^{-4}\).
Leave–one–out CV–MSE is slightly smaller for the logarithmic fit
(\(4.53\times10^{-4}\) vs.\ \(7.30\times10^{-4}\)), but the BIC differences in favor of the power law are decisive: \(\Delta\mathrm{BIC}_{\rm pow,log}\sim 41\). The best fit yields \(\alpha\simeq 0.30\) (robust across tail windows,
\(\alpha\in[0.22,0.48]\)), establishing \emph{sub–volume} growth of the measured entropy in the
heavy sector (see Table~\ref{tab:plateaus}).

\paragraph*{Light (\(\tau\)) leg — volume–law scaling.}
For the light leg [Fig.~\ref{fig:timeseries_dual}(b)], model selection based on BIC gives a tie between a linear and a power–law fit (both \(\mathrm{BIC}=-14.283\)), while leave–one–out CV–MSE favors the linear model (linear \(=1.532\times10^{-2}\) vs.\ power \(=2.457\times10^{-2}\)).
A linear fit \(S^{(\tau|\sigma)}_{\rm sat}(L)=aL+b\) yields
\(a=0.247876\) with a 95\% CI \([0.245651,\,0.250183]\) and \(b=-0.633417\).
The best power–law fit returns \(\alpha\simeq 1\), indistinguishable from linear by BIC, confirming extensive (thermal) \emph{volume–law} scaling \(S^{(\tau|
	\sigma)}_{\rm sat}\propto L\) for the measured entropy in the light sector (see Table~\ref{tab:plateaus}).

\paragraph*{Reversed QDL.}
In the canonical QDL scenario~\cite{Grover2014}, the heavy degrees of freedom thermalize and,
\emph{after measuring the heavy species}, the remaining (“disentangled”) species exhibits an
\emph{area law} for the measured entanglement (in 1D, \(S=\mathcal{O}(1)\)).
In finite-size, dynamical settings this can appear as weaker \emph{sub-volume} growth
(e.g., \(S\sim L^{\alpha}\) with \(0<\alpha<1\)).
In our ladder the \emph{measured} entropy shows the \emph{opposite} hierarchy:
\begin{itemize}\setlength{\itemsep}{2pt}
	\item Light (\(\tau\)) spins: \(S^{(\tau|\sigma)}_{\mathrm{sat}}(L)\propto L\) (volume law; thermal).
	\item Heavy (\(\sigma\)) spins: \(S^{(\sigma|\tau)}_{\mathrm{sat}}(L)\propto L^{\alpha}\) with \(\alpha<1\)
	(sub-volume; localized). Our best fit gives \(\alpha\simeq 0.30\)
	(robust across analysis windows), decisively favored over linear and logarithmic
	alternatives by BIC, with small RSS and competitive CV–MSE.
\end{itemize}

For completeness, we examine the dependence of the long-time saturation values of the measured half-cut entanglement entropy on the inter-leg coupling $J_z$ at fixed system size $L=8$. The saturation values are extracted using the terminal window averaging procedure defined in Eq.~(\ref{eq:Ssat-discrete}) over the interval $t\in[10^{11},10^{12}]$. As shown in Fig.~\ref{fig:timeseries_dual}(c), both measurement protocols yield comparable entropies for weak coupling ($J_z\ll1$), close to the thermal (Page) value expected for a half bipartition of this system, $S_{\mathrm{Page}}\simeq \ln(2^{L/2})-\tfrac{1}{2}\approx2.27$ for $L=8$. 
However, as $J_z$ increases beyond $1$, a distinct separation emerges: $S^{(\tau|\sigma)}_{L/2}$ remains relatively larger, while $S^{(\sigma|\tau)}_{L/2}$ is strongly suppressed.
This behavior is consistent with the finite-size scaling analysis presented above, indicating that the light ($\tau$) sector remains thermal with volume-law measured entanglement, while the heavy ($\sigma$) sector exhibits strongly reduced, sub-volume entanglement consistent with partial localization, as expected in the reversed quantum disentangled liquid regime.

\emph{Remark on admissible scalings.} Our system is a fixed–width (quasi–1D) ladder, so the relevant
“volume’’ scales as \(V\!\sim\!W L\) with constant width \(W= 2\). Locality and Page–like bounds imply
\(S(A)\le \ln\,(\,\!\dim\mathcal{H}_A)=(W L)\ln 2\), hence \(S_{\rm sat}= O(L)\).
Although a two–parameter power law can occasionally favor an exponent \(\alpha>1\) on three sizes,
such super–linear fits are super–extensive and unreliable for extrapolation; we therefore restrict
to \(\alpha\le 1\). For the light leg this collapses to \(\alpha\simeq 1\) (volume law), while the
heavy leg exhibits genuine sub–volume scaling with \(\alpha<1\).

\begin{table}[t]
	\centering
	\caption{Saturation values of the \emph{measured} entropy \(S_{\rm sat}(L)\) with SEM uncertainties,
		best model, and inferred scaling.}
	\label{tab:plateaus}
	\vspace{3pt}
	\begin{tabular}{l r c c c}
		\hline\hline
		Sector & $L$ & $S_{\rm sat}\,\pm\,\mathrm{SEM}$ & Best fit & Scaling \\
		\hline
		 & 4 & $0.422503 \pm 0.010370$ &    power        &               \\
		$\sigma$ (heavy) & 6 & $0.681881 \pm 0.003113$ &  ($\alpha\simeq0.30$) & sub-volume \\
		& 8 & $0.886123 \pm 0.003142$ &            &               \\
		\hline
		   & 4 & $0.468061 \pm 0.008992$ &            &               \\
		$\tau$ (light)& 6 & $0.838122 \pm 0.002404$ & linear     & volume        \\
		& 8 & $1.351117 \pm 0.001058$ &            &               \\
		\hline\hline
	\end{tabular}
\end{table}

\section{Effective Two-Rung Theory in the Strong Coupling Limit}
\label{app:eff_Hamiltonian}

To develop an effective theory in the strong coupling limit \( J_z \gg J, J' \), we begin by analyzing the Hamiltonian for a pair of neighboring rungs \( \ell, \ell+1 \). The full spin ladder consists of two chains: one made of \(\sigma\)-spins (upper leg) and the other of \(\tau\)-spins (lower leg), both being spin-\(\tfrac{1}{2}\) degrees of freedom Fig.~\ref{fig:ladder_rung}.

The leading contribution to the rung interaction is encapsulated by the unperturbed Hamiltonian
\begin{equation}
    H_0 = \frac{J_z}{2} \sum_{\ell=1}^{2} \tau_\ell^z \sigma_\ell^z,
\end{equation}
where the prefactor \( J_z/2 \) represents the interaction strength \emph{per rung}. Since each rung contains a single \(\tau^z \sigma^z\) interaction, the total energy per pair of rungs would scale with \(J_z\), and thus we include the factor of \(1/2\) to avoid overcounting when generalizing to the full ladder denoted by \( H_{zz} \) in Eq.~\ref{eq:H}

The eigenstates of \( H_0 \) on each rung organize into two energy-separated doublets:

\begin{align}
	\left| \gamma_\ell^+ \right\rangle &= \left| \downarrow \right\rangle_\tau \otimes \left| \uparrow \right\rangle_\sigma, &
	\left| \gamma_\ell^- \right\rangle &= \left| \uparrow \right\rangle_\tau \otimes \left| \downarrow \right\rangle_\sigma, \\
	\left| \varphi_\ell^+ \right\rangle &= \left| \uparrow \right\rangle_\tau \otimes \left| \uparrow \right\rangle_\sigma, &
	\left| \varphi_\ell^- \right\rangle &= \left| \downarrow \right\rangle_\tau \otimes \left| \downarrow \right\rangle_\sigma.
\end{align}
The states \( \left| \gamma_\ell^\pm \right\rangle \) span the low-energy sector with energy \( -J_z \), while the \( \left| \varphi_\ell^\pm \right\rangle \) reside in the high-energy sector with energy \( +J_z \).

Within the doublet subspace \( \{ \left| \gamma_l^\pm \right\rangle \} \), we define pseudospin operators:
\begin{align}
	\hat{\Gamma}_l^+ &= \left| \gamma_l^+ \right\rangle \left\langle \gamma_l^- \right|, \notag\\
	\hat{\Gamma}_l^- &= \left| \gamma_l^- \right\rangle \left\langle \gamma_l^+ \right|, \notag\\
	\hat{\Gamma}_l^z &= \left| \gamma_l^+ \right\rangle \left\langle \gamma_l^+ \right| - \left| \gamma_l^- \right\rangle \left\langle \gamma_l^- \right|.
\end{align}

We introduce the perturbing terms:
\begin{align}
	V_1 &= 2J \left( \tau_\ell^+ \tau_{\ell+1}^- + \tau_\ell^- \tau_{\ell+1}^+ \right), \\
	V_2 &= 2J' \left( \sigma_\ell^+ \sigma_{\ell+1}^- + \sigma_\ell^- \sigma_{\ell+1}^+ \right),
\end{align}
with total perturbation \( V = V_1 + V_2 \).

The spectral decomposition of \( H_0 \) implies a two-level structure and a clear energy gap of \( 2J_z \). Projecting onto the low-energy subspace using:
\begin{figure}[t]
	\centering
	\includegraphics[width=1.05\linewidth, trim=120 230 120 90, clip]{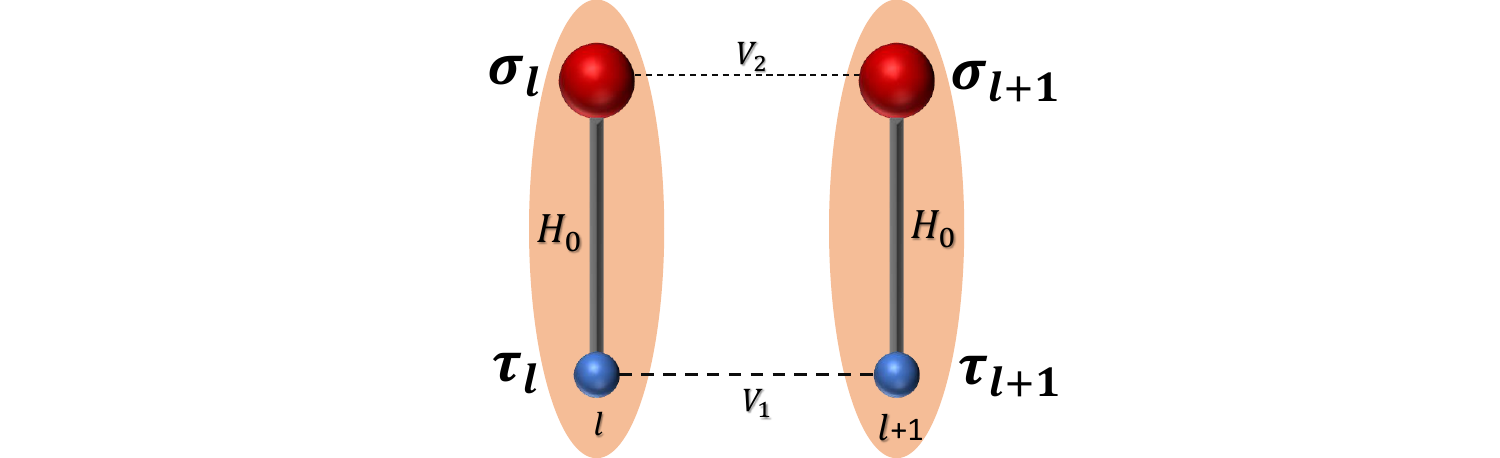}
	\caption{%
		(Color online) Schematic of two rungs of the spin ladder. The dominant rung coupling \( H_0 \) (solid lines) defines the unperturbed Hamiltonian, while the leg couplings \( V_1 \) and \( V_2 \) (dashed lines) are treated perturbatively. Each rung hosts a pair of coupled spins \( (\tau_l, \sigma_l) \), forming the basis for the effective low- and high-energy subspaces.
	}
	\label{fig:ladder_rung}
\end{figure}
\begin{align}
	P &= P_\ell \otimes P_{\ell+1}, \quad
	Q = \mathbb{1} - P, \\
	P_\ell &= \sum_{\pm} \left| \gamma_\ell^\pm \right\rangle \left\langle \gamma_\ell^\pm \right|, \quad
	Q_\ell = \sum_{\pm} \left| \varphi_\ell^\pm \right\rangle \left\langle \varphi_\ell^\pm \right|,
\end{align}
we compute the second-order contribution to the effective Hamiltonian using Brillouin--Wigner perturbation theory:
\begin{equation}
	H_\text{eff}^{(\text{low})} = -\frac{1}{2 J_z} P V Q V P.
\end{equation}
Evaluating each term:
\begin{align}
	P V_1 Q V_1 P &= 2J^2 \left( \mathbb{1} - \Gamma_\ell^z \Gamma_{\ell+1}^z \right), \\
	P V_2 Q V_2 P &= 2{J'}^2 \left( \mathbb{1} - \Gamma_\ell^z \Gamma_{\ell+1}^z \right), \\
	P V_2 Q V_1 P &= 2JJ' \left( \Gamma_\ell^x \Gamma_{\ell+1}^x + \Gamma_\ell^y \Gamma_{\ell+1}^y \right), \\
	P V_1 Q V_2 P &= 2JJ' \left( \Gamma_\ell^x \Gamma_{\ell+1}^x + \Gamma_\ell^y \Gamma_{\ell+1}^y \right),
\end{align}
the final effective Hamiltonian reads:
\begin{align}
	\mathcal{H}_{\text{eff}}^{\text{(low)}} =\ 
	& -\frac{2 J J'}{J_z}
	\left(
	\Gamma_\ell^x \Gamma_{\ell+1}^x + \Gamma_\ell^y \Gamma_{\ell+1}^y
	\right) \notag \\
	& + \left(\frac{J^2 + J'^2}{J_z}\right) \Gamma_\ell^z \Gamma_{\ell+1}^z
	- \left( \frac{J^2 + J'^2}{J_z} \right) \mathbb{1}.
\end{align}


\begin{thebibliography}{74}%
\makeatletter
\providecommand \@ifxundefined [1]{%
 \@ifx{#1\undefined}
}%
\providecommand \@ifnum [1]{%
 \ifnum #1\expandafter \@firstoftwo
 \else \expandafter \@secondoftwo
 \fi
}%
\providecommand \@ifx [1]{%
 \ifx #1\expandafter \@firstoftwo
 \else \expandafter \@secondoftwo
 \fi
}%
\providecommand \natexlab [1]{#1}%
\providecommand \enquote  [1]{``#1''}%
\providecommand \bibnamefont  [1]{#1}%
\providecommand \bibfnamefont [1]{#1}%
\providecommand \citenamefont [1]{#1}%
\providecommand \href@noop [0]{\@secondoftwo}%
\providecommand \href [0]{\begingroup \@sanitize@url \@href}%
\providecommand \@href[1]{\@@startlink{#1}\@@href}%
\providecommand \@@href[1]{\endgroup#1\@@endlink}%
\providecommand \@sanitize@url [0]{\catcode `\\12\catcode `\$12\catcode
  `\&12\catcode `\#12\catcode `\^12\catcode `\_12\catcode `\%12\relax}%
\providecommand \@@startlink[1]{}%
\providecommand \@@endlink[0]{}%
\providecommand \url  [0]{\begingroup\@sanitize@url \@url }%
\providecommand \@url [1]{\endgroup\@href {#1}{\urlprefix }}%
\providecommand \urlprefix  [0]{URL }%
\providecommand \Eprint [0]{\href }%
\providecommand \doibase [0]{https://doi.org/}%
\providecommand \selectlanguage [0]{\@gobble}%
\providecommand \bibinfo  [0]{\@secondoftwo}%
\providecommand \bibfield  [0]{\@secondoftwo}%
\providecommand \translation [1]{[#1]}%
\providecommand \BibitemOpen [0]{}%
\providecommand \bibitemStop [0]{}%
\providecommand \bibitemNoStop [0]{.\EOS\space}%
\providecommand \EOS [0]{\spacefactor3000\relax}%
\providecommand \BibitemShut  [1]{\csname bibitem#1\endcsname}%
\let\auto@bib@innerbib\@empty
\bibitem [{\citenamefont {Deutsch}(1991)}]{Deutsch1991}%
  \BibitemOpen
  \bibfield  {author} {\bibinfo {author} {\bibfnamefont {J.~M.}\ \bibnamefont
  {Deutsch}},\ }\bibfield  {title} {\bibinfo {title} {Quantum statistical
  mechanics in a closed system},\ }\href
  {https://doi.org/10.1103/PhysRevA.43.2046} {\bibfield  {journal} {\bibinfo
  {journal} {Phys. Rev. A}\ }\textbf {\bibinfo {volume} {43}},\ \bibinfo
  {pages} {2046} (\bibinfo {year} {1991})}\BibitemShut {NoStop}%
\bibitem [{\citenamefont {Srednicki}(1994)}]{Srednicki1994}%
  \BibitemOpen
  \bibfield  {author} {\bibinfo {author} {\bibfnamefont {M.}~\bibnamefont
  {Srednicki}},\ }\bibfield  {title} {\bibinfo {title} {Chaos and quantum
  thermalization},\ }\href {https://doi.org/10.1103/PhysRevE.50.888} {\bibfield
   {journal} {\bibinfo  {journal} {Phys. Rev. E}\ }\textbf {\bibinfo {volume}
  {50}},\ \bibinfo {pages} {888} (\bibinfo {year} {1994})}\BibitemShut
  {NoStop}%
\bibitem [{\citenamefont {Rigol}\ \emph {et~al.}(2008)\citenamefont {Rigol},
  \citenamefont {Dunjko},\ and\ \citenamefont {Olshanii}}]{Rigol2008}%
  \BibitemOpen
  \bibfield  {author} {\bibinfo {author} {\bibfnamefont {M.}~\bibnamefont
  {Rigol}}, \bibinfo {author} {\bibfnamefont {V.}~\bibnamefont {Dunjko}},\ and\
  \bibinfo {author} {\bibfnamefont {M.}~\bibnamefont {Olshanii}},\ }\bibfield
  {title} {\bibinfo {title} {Thermalization and its mechanism for generic
  isolated quantum systems},\ }\href {https://doi.org/10.1038/nature06838}
  {\bibfield  {journal} {\bibinfo  {journal} {Nature}\ }\textbf {\bibinfo
  {volume} {452}},\ \bibinfo {pages} {854–858} (\bibinfo {year}
  {2008})}\BibitemShut {NoStop}%
\bibitem [{\citenamefont {D’Alessio}\ \emph {et~al.}(2016)\citenamefont
  {D’Alessio}, \citenamefont {Kafri}, \citenamefont {Polkovnikov},\ and\
  \citenamefont {Rigol}}]{DAlessio2016}%
  \BibitemOpen
  \bibfield  {author} {\bibinfo {author} {\bibfnamefont {L.}~\bibnamefont
  {D’Alessio}}, \bibinfo {author} {\bibfnamefont {Y.}~\bibnamefont {Kafri}},
  \bibinfo {author} {\bibfnamefont {A.}~\bibnamefont {Polkovnikov}},\ and\
  \bibinfo {author} {\bibfnamefont {M.}~\bibnamefont {Rigol}},\ }\bibfield
  {title} {\bibinfo {title} {From quantum chaos and eigenstate thermalization
  to statistical mechanics and thermodynamics},\ }\href
  {https://doi.org/10.1080/00018732.2016.1198134} {\bibfield  {journal}
  {\bibinfo  {journal} {Advances in Physics}\ }\textbf {\bibinfo {volume}
  {65}},\ \bibinfo {pages} {239–362} (\bibinfo {year} {2016})}\BibitemShut
  {NoStop}%
\bibitem [{\citenamefont {Basko}\ \emph {et~al.}(2006)\citenamefont {Basko},
  \citenamefont {Aleiner},\ and\ \citenamefont {Altshuler}}]{Basko2006}%
  \BibitemOpen
  \bibfield  {author} {\bibinfo {author} {\bibfnamefont {D.}~\bibnamefont
  {Basko}}, \bibinfo {author} {\bibfnamefont {I.}~\bibnamefont {Aleiner}},\
  and\ \bibinfo {author} {\bibfnamefont {B.}~\bibnamefont {Altshuler}},\
  }\bibfield  {title} {\bibinfo {title} {Metal–insulator transition in a
  weakly interacting many-electron system with localized single-particle
  states},\ }\href {https://doi.org/10.1016/j.aop.2005.11.014} {\bibfield
  {journal} {\bibinfo  {journal} {Annals of Physics}\ }\textbf {\bibinfo
  {volume} {321}},\ \bibinfo {pages} {1126–1205} (\bibinfo {year}
  {2006})}\BibitemShut {NoStop}%
\bibitem [{\citenamefont {Oganesyan}\ and\ \citenamefont
  {Huse}(2007)}]{Oganesyan2007}%
  \BibitemOpen
  \bibfield  {author} {\bibinfo {author} {\bibfnamefont {V.}~\bibnamefont
  {Oganesyan}}\ and\ \bibinfo {author} {\bibfnamefont {D.~A.}\ \bibnamefont
  {Huse}},\ }\bibfield  {title} {\bibinfo {title} {Localization of interacting
  fermions at high temperature},\ }\bibfield  {journal} {\bibinfo  {journal}
  {Physical Review B}\ }\textbf {\bibinfo {volume} {75}},\ \href
  {https://doi.org/10.1103/physrevb.75.155111} {10.1103/physrevb.75.155111}
  (\bibinfo {year} {2007})\BibitemShut {NoStop}%
\bibitem [{\citenamefont {Nandkishore}\ and\ \citenamefont
  {Huse}(2015)}]{Nandkishore2015}%
  \BibitemOpen
  \bibfield  {author} {\bibinfo {author} {\bibfnamefont {R.}~\bibnamefont
  {Nandkishore}}\ and\ \bibinfo {author} {\bibfnamefont {D.~A.}\ \bibnamefont
  {Huse}},\ }\bibfield  {title} {\bibinfo {title} {Many-body localization and
  thermalization in quantum statistical mechanics},\ }\href
  {https://doi.org/10.1146/annurev-conmatphys-031214-014726} {\bibfield
  {journal} {\bibinfo  {journal} {Annual Review of Condensed Matter Physics}\
  }\textbf {\bibinfo {volume} {6}},\ \bibinfo {pages} {15–38} (\bibinfo
  {year} {2015})}\BibitemShut {NoStop}%
\bibitem [{\citenamefont {Sierant}\ \emph {et~al.}(2025)\citenamefont
  {Sierant}, \citenamefont {Lewenstein}, \citenamefont {Scardicchio},
  \citenamefont {Vidmar},\ and\ \citenamefont {Zakrzewski}}]{Sierant2025}%
  \BibitemOpen
  \bibfield  {author} {\bibinfo {author} {\bibfnamefont {P.}~\bibnamefont
  {Sierant}}, \bibinfo {author} {\bibfnamefont {M.}~\bibnamefont {Lewenstein}},
  \bibinfo {author} {\bibfnamefont {A.}~\bibnamefont {Scardicchio}}, \bibinfo
  {author} {\bibfnamefont {L.}~\bibnamefont {Vidmar}},\ and\ \bibinfo {author}
  {\bibfnamefont {J.}~\bibnamefont {Zakrzewski}},\ }\bibfield  {title}
  {\bibinfo {title} {Many-body localization in the age of classical
  computing*},\ }\href {https://doi.org/10.1088/1361-6633/ad9756} {\bibfield
  {journal} {\bibinfo  {journal} {Reports on Progress in Physics}\ }\textbf
  {\bibinfo {volume} {88}},\ \bibinfo {pages} {026502} (\bibinfo {year}
  {2025})}\BibitemShut {NoStop}%
\bibitem [{\citenamefont {Abanin}\ \emph {et~al.}(2019)\citenamefont {Abanin},
  \citenamefont {Altman}, \citenamefont {Bloch},\ and\ \citenamefont
  {Serbyn}}]{Abanin2019}%
  \BibitemOpen
  \bibfield  {author} {\bibinfo {author} {\bibfnamefont {D.~A.}\ \bibnamefont
  {Abanin}}, \bibinfo {author} {\bibfnamefont {E.}~\bibnamefont {Altman}},
  \bibinfo {author} {\bibfnamefont {I.}~\bibnamefont {Bloch}},\ and\ \bibinfo
  {author} {\bibfnamefont {M.}~\bibnamefont {Serbyn}},\ }\bibfield  {title}
  {\bibinfo {title} {Colloquium: Many-body localization, thermalization, and
  entanglement},\ }\href {https://doi.org/10.1103/RevModPhys.91.021001}
  {\bibfield  {journal} {\bibinfo  {journal} {Rev. Mod. Phys.}\ }\textbf
  {\bibinfo {volume} {91}},\ \bibinfo {pages} {021001} (\bibinfo {year}
  {2019})}\BibitemShut {NoStop}%
\bibitem [{\citenamefont {Bu\ifmmode~\check{c}\else
  \v{c}\fi{}a}(2023)}]{Buca2023}%
  \BibitemOpen
  \bibfield  {author} {\bibinfo {author} {\bibfnamefont {B.}~\bibnamefont
  {Bu\ifmmode~\check{c}\else \v{c}\fi{}a}},\ }\bibfield  {title} {\bibinfo
  {title} {Unified theory of local quantum many-body dynamics: Eigenoperator
  thermalization theorems},\ }\href
  {https://doi.org/10.1103/PhysRevX.13.031013} {\bibfield  {journal} {\bibinfo
  {journal} {Phys. Rev. X}\ }\textbf {\bibinfo {volume} {13}},\ \bibinfo
  {pages} {031013} (\bibinfo {year} {2023})}\BibitemShut {NoStop}%
\bibitem [{\citenamefont {Sala}\ \emph {et~al.}(2020)\citenamefont {Sala},
  \citenamefont {Rakovszky}, \citenamefont {Verresen}, \citenamefont {Knap},\
  and\ \citenamefont {Pollmann}}]{Sala2020}%
  \BibitemOpen
  \bibfield  {author} {\bibinfo {author} {\bibfnamefont {P.}~\bibnamefont
  {Sala}}, \bibinfo {author} {\bibfnamefont {T.}~\bibnamefont {Rakovszky}},
  \bibinfo {author} {\bibfnamefont {R.}~\bibnamefont {Verresen}}, \bibinfo
  {author} {\bibfnamefont {M.}~\bibnamefont {Knap}},\ and\ \bibinfo {author}
  {\bibfnamefont {F.}~\bibnamefont {Pollmann}},\ }\bibfield  {title} {\bibinfo
  {title} {Ergodicity breaking arising from hilbert space fragmentation in
  dipole-conserving hamiltonians},\ }\bibfield  {journal} {\bibinfo  {journal}
  {Physical Review X}\ }\textbf {\bibinfo {volume} {10}},\ \href
  {https://doi.org/10.1103/physrevx.10.011047} {10.1103/physrevx.10.011047}
  (\bibinfo {year} {2020})\BibitemShut {NoStop}%
\bibitem [{\citenamefont {Khemani}\ \emph {et~al.}(2020)\citenamefont
  {Khemani}, \citenamefont {Hermele},\ and\ \citenamefont
  {Nandkishore}}]{Khemani2020}%
  \BibitemOpen
  \bibfield  {author} {\bibinfo {author} {\bibfnamefont {V.}~\bibnamefont
  {Khemani}}, \bibinfo {author} {\bibfnamefont {M.}~\bibnamefont {Hermele}},\
  and\ \bibinfo {author} {\bibfnamefont {R.}~\bibnamefont {Nandkishore}},\
  }\bibfield  {title} {\bibinfo {title} {Localization from hilbert space
  shattering: From theory to physical realizations},\ }\bibfield  {journal}
  {\bibinfo  {journal} {Physical Review B}\ }\textbf {\bibinfo {volume}
  {101}},\ \href {https://doi.org/10.1103/physrevb.101.174204}
  {10.1103/physrevb.101.174204} (\bibinfo {year} {2020})\BibitemShut {NoStop}%
\bibitem [{\citenamefont {Jeyaretnam}\ \emph {et~al.}(2025)\citenamefont
  {Jeyaretnam}, \citenamefont {Bhore}, \citenamefont {Osborne}, \citenamefont
  {Halimeh},\ and\ \citenamefont {Papić}}]{Jeyaretnam2025}%
  \BibitemOpen
  \bibfield  {author} {\bibinfo {author} {\bibfnamefont {J.}~\bibnamefont
  {Jeyaretnam}}, \bibinfo {author} {\bibfnamefont {T.}~\bibnamefont {Bhore}},
  \bibinfo {author} {\bibfnamefont {J.~J.}\ \bibnamefont {Osborne}}, \bibinfo
  {author} {\bibfnamefont {J.~C.}\ \bibnamefont {Halimeh}},\ and\ \bibinfo
  {author} {\bibfnamefont {Z.}~\bibnamefont {Papić}},\ }\bibfield  {title}
  {\bibinfo {title} {Hilbert space fragmentation at the origin of disorder-free
  localization in the lattice schwinger model},\ }\bibfield  {journal}
  {\bibinfo  {journal} {Communications Physics}\ }\textbf {\bibinfo {volume}
  {8}},\ \href {https://doi.org/10.1038/s42005-025-02039-8}
  {10.1038/s42005-025-02039-8} (\bibinfo {year} {2025})\BibitemShut {NoStop}%
\bibitem [{\citenamefont {Yang}\ \emph {et~al.}(2025)\citenamefont {Yang},
  \citenamefont {Yarloo}, \citenamefont {Zhang}, \citenamefont {Mølmer},\ and\
  \citenamefont {Nielsen}}]{Yang2025}%
  \BibitemOpen
  \bibfield  {author} {\bibinfo {author} {\bibfnamefont {F.}~\bibnamefont
  {Yang}}, \bibinfo {author} {\bibfnamefont {H.}~\bibnamefont {Yarloo}},
  \bibinfo {author} {\bibfnamefont {H.-C.}\ \bibnamefont {Zhang}}, \bibinfo
  {author} {\bibfnamefont {K.}~\bibnamefont {Mølmer}},\ and\ \bibinfo {author}
  {\bibfnamefont {A.~E.~B.}\ \bibnamefont {Nielsen}},\ }\bibfield  {title}
  {\bibinfo {title} {Probing hilbert space fragmentation with strongly
  interacting rydberg atoms},\ }\bibfield  {journal} {\bibinfo  {journal}
  {Physical Review B}\ }\textbf {\bibinfo {volume} {111}},\ \href
  {https://doi.org/10.1103/physrevb.111.144313} {10.1103/physrevb.111.144313}
  (\bibinfo {year} {2025})\BibitemShut {NoStop}%
\bibitem [{\citenamefont {Turner}\ \emph {et~al.}(2018)\citenamefont {Turner},
  \citenamefont {Michailidis}, \citenamefont {Abanin}, \citenamefont {Serbyn},\
  and\ \citenamefont {Papić}}]{Turner2018}%
  \BibitemOpen
  \bibfield  {author} {\bibinfo {author} {\bibfnamefont {C.~J.}\ \bibnamefont
  {Turner}}, \bibinfo {author} {\bibfnamefont {A.~A.}\ \bibnamefont
  {Michailidis}}, \bibinfo {author} {\bibfnamefont {D.~A.}\ \bibnamefont
  {Abanin}}, \bibinfo {author} {\bibfnamefont {M.}~\bibnamefont {Serbyn}},\
  and\ \bibinfo {author} {\bibfnamefont {Z.}~\bibnamefont {Papić}},\
  }\bibfield  {title} {\bibinfo {title} {Weak ergodicity breaking from quantum
  many-body scars},\ }\href@noop {} {\bibfield  {journal} {\bibinfo  {journal}
  {Nat. Phys.}\ }\textbf {\bibinfo {volume} {14}},\ \bibinfo {pages} {745}
  (\bibinfo {year} {2018})}\BibitemShut {NoStop}%
\bibitem [{\citenamefont {Serbyn}\ \emph {et~al.}(2021)\citenamefont {Serbyn},
  \citenamefont {Abanin},\ and\ \citenamefont {Papić}}]{Serbyn2021}%
  \BibitemOpen
  \bibfield  {author} {\bibinfo {author} {\bibfnamefont {M.}~\bibnamefont
  {Serbyn}}, \bibinfo {author} {\bibfnamefont {D.~A.}\ \bibnamefont {Abanin}},\
  and\ \bibinfo {author} {\bibfnamefont {Z.}~\bibnamefont {Papić}},\
  }\bibfield  {title} {\bibinfo {title} {Quantum many-body scars and weak
  breaking of ergodicity},\ }\href {https://doi.org/10.1038/s41567-021-01230-2}
  {\bibfield  {journal} {\bibinfo  {journal} {Nature Physics}\ }\textbf
  {\bibinfo {volume} {17}},\ \bibinfo {pages} {675–685} (\bibinfo {year}
  {2021})}\BibitemShut {NoStop}%
\bibitem [{\citenamefont {Srivatsa}\ \emph {et~al.}(2023)\citenamefont
  {Srivatsa}, \citenamefont {Yarloo}, \citenamefont {Moessner},\ and\
  \citenamefont {Nielsen}}]{Srivatsa2023}%
  \BibitemOpen
  \bibfield  {author} {\bibinfo {author} {\bibfnamefont {N.~S.}\ \bibnamefont
  {Srivatsa}}, \bibinfo {author} {\bibfnamefont {H.}~\bibnamefont {Yarloo}},
  \bibinfo {author} {\bibfnamefont {R.}~\bibnamefont {Moessner}},\ and\
  \bibinfo {author} {\bibfnamefont {A.~E.~B.}\ \bibnamefont {Nielsen}},\
  }\bibfield  {title} {\bibinfo {title} {Mobility edges through inverted
  quantum many-body scarring},\ }\bibfield  {journal} {\bibinfo  {journal}
  {Physical Review B}\ }\textbf {\bibinfo {volume} {108}},\ \href
  {https://doi.org/10.1103/physrevb.108.l100202} {10.1103/physrevb.108.l100202}
  (\bibinfo {year} {2023})\BibitemShut {NoStop}%
\bibitem [{\citenamefont {Yarloo}\ \emph {et~al.}(2024)\citenamefont {Yarloo},
  \citenamefont {Zhang},\ and\ \citenamefont {Nielsen}}]{Yarloo2024}%
  \BibitemOpen
  \bibfield  {author} {\bibinfo {author} {\bibfnamefont {H.}~\bibnamefont
  {Yarloo}}, \bibinfo {author} {\bibfnamefont {H.-C.}\ \bibnamefont {Zhang}},\
  and\ \bibinfo {author} {\bibfnamefont {A.~E.~B.}\ \bibnamefont {Nielsen}},\
  }\bibfield  {title} {\bibinfo {title} {Adiabatic time evolution of highly
  excited states},\ }\href {https://doi.org/10.1103/PRXQuantum.5.020365}
  {\bibfield  {journal} {\bibinfo  {journal} {PRX Quantum}\ }\textbf {\bibinfo
  {volume} {5}},\ \bibinfo {pages} {020365} (\bibinfo {year} {2024})},\ \Eprint
  {https://arxiv.org/abs/2306.13967} {arXiv:2306.13967 [quant-ph]} \BibitemShut
  {NoStop}%
\bibitem [{\citenamefont {Michailidis}\ \emph {et~al.}(2020)\citenamefont
  {Michailidis}, \citenamefont {Turner}, \citenamefont
  {Papi\ifmmode~\acute{c}\else \'{c}\fi{}}, \citenamefont {Abanin},\ and\
  \citenamefont {Serbyn}}]{Michailidis2020}%
  \BibitemOpen
  \bibfield  {author} {\bibinfo {author} {\bibfnamefont {A.~A.}\ \bibnamefont
  {Michailidis}}, \bibinfo {author} {\bibfnamefont {C.~J.}\ \bibnamefont
  {Turner}}, \bibinfo {author} {\bibfnamefont {Z.}~\bibnamefont
  {Papi\ifmmode~\acute{c}\else \'{c}\fi{}}}, \bibinfo {author} {\bibfnamefont
  {D.~A.}\ \bibnamefont {Abanin}},\ and\ \bibinfo {author} {\bibfnamefont
  {M.}~\bibnamefont {Serbyn}},\ }\bibfield  {title} {\bibinfo {title} {Slow
  quantum thermalization and many-body revivals from mixed phase space},\
  }\href {https://doi.org/10.1103/PhysRevX.10.011055} {\bibfield  {journal}
  {\bibinfo  {journal} {Phys. Rev. X}\ }\textbf {\bibinfo {volume} {10}},\
  \bibinfo {pages} {011055} (\bibinfo {year} {2020})}\BibitemShut {NoStop}%
\bibitem [{\citenamefont {van Nieuwenburg}\ \emph {et~al.}(2019)\citenamefont
  {van Nieuwenburg}, \citenamefont {Baum},\ and\ \citenamefont
  {Refael}}]{Evert2019}%
  \BibitemOpen
  \bibfield  {author} {\bibinfo {author} {\bibfnamefont {E.}~\bibnamefont {van
  Nieuwenburg}}, \bibinfo {author} {\bibfnamefont {Y.}~\bibnamefont {Baum}},\
  and\ \bibinfo {author} {\bibfnamefont {G.}~\bibnamefont {Refael}},\
  }\bibfield  {title} {\bibinfo {title} {From bloch oscillations to many-body
  localization in clean interacting systems},\ }\href
  {https://doi.org/10.1073/pnas.1819316116} {\bibfield  {journal} {\bibinfo
  {journal} {Proceedings of the National Academy of Sciences}\ }\textbf
  {\bibinfo {volume} {116}},\ \bibinfo {pages} {9269} (\bibinfo {year}
  {2019})},\ \Eprint
  {https://arxiv.org/abs/https://www.pnas.org/doi/pdf/10.1073/pnas.1819316116}
  {https://www.pnas.org/doi/pdf/10.1073/pnas.1819316116} \BibitemShut {NoStop}%
\bibitem [{\citenamefont {Schulz}\ \emph {et~al.}(2019)\citenamefont {Schulz},
  \citenamefont {Hooley}, \citenamefont {Moessner},\ and\ \citenamefont
  {Pollmann}}]{Schulz2019}%
  \BibitemOpen
  \bibfield  {author} {\bibinfo {author} {\bibfnamefont {M.}~\bibnamefont
  {Schulz}}, \bibinfo {author} {\bibfnamefont {C.~A.}\ \bibnamefont {Hooley}},
  \bibinfo {author} {\bibfnamefont {R.}~\bibnamefont {Moessner}},\ and\
  \bibinfo {author} {\bibfnamefont {F.}~\bibnamefont {Pollmann}},\ }\bibfield
  {title} {\bibinfo {title} {Stark many-body localization},\ }\href
  {https://doi.org/10.1103/PhysRevLett.122.040606} {\bibfield  {journal}
  {\bibinfo  {journal} {Phys. Rev. Lett.}\ }\textbf {\bibinfo {volume} {122}},\
  \bibinfo {pages} {040606} (\bibinfo {year} {2019})}\BibitemShut {NoStop}%
\bibitem [{\citenamefont {Grover}\ and\ \citenamefont
  {Fisher}(2014)}]{Grover2014}%
  \BibitemOpen
  \bibfield  {author} {\bibinfo {author} {\bibfnamefont {T.}~\bibnamefont
  {Grover}}\ and\ \bibinfo {author} {\bibfnamefont {M.~P.~A.}\ \bibnamefont
  {Fisher}},\ }\bibfield  {title} {\bibinfo {title} {Quantum disentangled
  liquids},\ }\href {https://doi.org/10.1088/1742-5468/2014/10/p10010}
  {\bibfield  {journal} {\bibinfo  {journal} {Journal of Statistical Mechanics:
  Theory and Experiment}\ }\textbf {\bibinfo {volume} {2014}},\ \bibinfo
  {pages} {P10010} (\bibinfo {year} {2014})}\BibitemShut {NoStop}%
\bibitem [{\citenamefont {Veness}\ \emph {et~al.}(2017)\citenamefont {Veness},
  \citenamefont {Essler},\ and\ \citenamefont {Fisher}}]{Veness2017}%
  \BibitemOpen
  \bibfield  {author} {\bibinfo {author} {\bibfnamefont {T.}~\bibnamefont
  {Veness}}, \bibinfo {author} {\bibfnamefont {F.~H.~L.}\ \bibnamefont
  {Essler}},\ and\ \bibinfo {author} {\bibfnamefont {M.~P.~A.}\ \bibnamefont
  {Fisher}},\ }\bibfield  {title} {\bibinfo {title} {Quantum disentangled
  liquid in the half-filled hubbard model},\ }\bibfield  {journal} {\bibinfo
  {journal} {Physical Review B}\ }\textbf {\bibinfo {volume} {96}},\ \href
  {https://doi.org/10.1103/physrevb.96.195153} {10.1103/physrevb.96.195153}
  (\bibinfo {year} {2017})\BibitemShut {NoStop}%
\bibitem [{\citenamefont {Abbasgholinejad}\ \emph {et~al.}(2023)\citenamefont
  {Abbasgholinejad}, \citenamefont {Raeisi},\ and\ \citenamefont
  {Langari}}]{Abbasgholinejad2023}%
  \BibitemOpen
  \bibfield  {author} {\bibinfo {author} {\bibfnamefont {E.}~\bibnamefont
  {Abbasgholinejad}}, \bibinfo {author} {\bibfnamefont {S.}~\bibnamefont
  {Raeisi}},\ and\ \bibinfo {author} {\bibfnamefont {A.}~\bibnamefont
  {Langari}},\ }\bibfield  {title} {\bibinfo {title} {Numerical and quantum
  simulation of a quantum disentangled liquid},\ }\href
  {https://doi.org/10.1016/j.physa.2023.128561} {\bibfield  {journal} {\bibinfo
   {journal} {Physica A: Statistical Mechanics and its Applications}\ }\textbf
  {\bibinfo {volume} {615}},\ \bibinfo {pages} {128561} (\bibinfo {year}
  {2023})}\BibitemShut {NoStop}%
\bibitem [{\citenamefont {Smith}\ \emph {et~al.}(2017)\citenamefont {Smith},
  \citenamefont {Knolle}, \citenamefont {Moessner},\ and\ \citenamefont
  {Kovrizhin}}]{Smith2017}%
  \BibitemOpen
  \bibfield  {author} {\bibinfo {author} {\bibfnamefont {A.}~\bibnamefont
  {Smith}}, \bibinfo {author} {\bibfnamefont {J.}~\bibnamefont {Knolle}},
  \bibinfo {author} {\bibfnamefont {R.}~\bibnamefont {Moessner}},\ and\
  \bibinfo {author} {\bibfnamefont {D.}~\bibnamefont {Kovrizhin}},\ }\bibfield
  {title} {\bibinfo {title} {Absence of ergodicity without quenched disorder:
  From quantum disentangled liquids to many-body localization},\ }\bibfield
  {journal} {\bibinfo  {journal} {Physical Review Letters}\ }\textbf {\bibinfo
  {volume} {119}},\ \href {https://doi.org/10.1103/physrevlett.119.176601}
  {10.1103/physrevlett.119.176601} (\bibinfo {year} {2017})\BibitemShut
  {NoStop}%
\bibitem [{\citenamefont {Yao}\ \emph {et~al.}(2016)\citenamefont {Yao},
  \citenamefont {Laumann}, \citenamefont {Cirac}, \citenamefont {Lukin},\ and\
  \citenamefont {Moore}}]{Yao2016}%
  \BibitemOpen
  \bibfield  {author} {\bibinfo {author} {\bibfnamefont {N.}~\bibnamefont
  {Yao}}, \bibinfo {author} {\bibfnamefont {C.}~\bibnamefont {Laumann}},
  \bibinfo {author} {\bibfnamefont {J.}~\bibnamefont {Cirac}}, \bibinfo
  {author} {\bibfnamefont {M.}~\bibnamefont {Lukin}},\ and\ \bibinfo {author}
  {\bibfnamefont {J.}~\bibnamefont {Moore}},\ }\bibfield  {title} {\bibinfo
  {title} {Quasi-many-body localization in translation-invariant systems},\
  }\bibfield  {journal} {\bibinfo  {journal} {Physical Review Letters}\
  }\textbf {\bibinfo {volume} {117}},\ \href
  {https://doi.org/10.1103/physrevlett.117.240601}
  {10.1103/physrevlett.117.240601} (\bibinfo {year} {2016})\BibitemShut
  {NoStop}%
\bibitem [{\citenamefont {Yarloo}\ \emph {et~al.}(2018)\citenamefont {Yarloo},
  \citenamefont {Langari},\ and\ \citenamefont {Vaezi}}]{Yarloo2018}%
  \BibitemOpen
  \bibfield  {author} {\bibinfo {author} {\bibfnamefont {H.}~\bibnamefont
  {Yarloo}}, \bibinfo {author} {\bibfnamefont {A.}~\bibnamefont {Langari}},\
  and\ \bibinfo {author} {\bibfnamefont {A.}~\bibnamefont {Vaezi}},\ }\bibfield
   {title} {\bibinfo {title} {Anyonic self-induced disorder in a stabilizer
  code: Quasi many-body localization in a translational invariant model},\
  }\bibfield  {journal} {\bibinfo  {journal} {Physical Review B}\ }\textbf
  {\bibinfo {volume} {97}},\ \href {https://doi.org/10.1103/physrevb.97.054304}
  {10.1103/physrevb.97.054304} (\bibinfo {year} {2018})\BibitemShut {NoStop}%
\bibitem [{\citenamefont {Sala}\ \emph {et~al.}(2024)\citenamefont {Sala},
  \citenamefont {Giudici},\ and\ \citenamefont {Halimeh}}]{Sala2024}%
  \BibitemOpen
  \bibfield  {author} {\bibinfo {author} {\bibfnamefont {P.}~\bibnamefont
  {Sala}}, \bibinfo {author} {\bibfnamefont {G.}~\bibnamefont {Giudici}},\ and\
  \bibinfo {author} {\bibfnamefont {J.~C.}\ \bibnamefont {Halimeh}},\
  }\bibfield  {title} {\bibinfo {title} {Disorder-free localization as a purely
  classical effect},\ }\bibfield  {journal} {\bibinfo  {journal} {Physical
  Review B}\ }\textbf {\bibinfo {volume} {109}},\ \href
  {https://doi.org/10.1103/physrevb.109.l060305} {10.1103/physrevb.109.l060305}
  (\bibinfo {year} {2024})\BibitemShut {NoStop}%
\bibitem [{\citenamefont {Gunawardana}\ and\ \citenamefont
  {Uchoa}(2024)}]{Gunawardana2024}%
  \BibitemOpen
  \bibfield  {author} {\bibinfo {author} {\bibfnamefont {K.~G. S.~H.}\
  \bibnamefont {Gunawardana}}\ and\ \bibinfo {author} {\bibfnamefont
  {B.}~\bibnamefont {Uchoa}},\ }\bibfield  {title} {\bibinfo {title} {Disorder
  free many-body localization transition in two quasiperiodically coupled
  heisenberg spin chains},\ }\bibfield  {journal} {\bibinfo  {journal}
  {Physical Review B}\ }\textbf {\bibinfo {volume} {110}},\ \href
  {https://doi.org/10.1103/physrevb.110.054207} {10.1103/physrevb.110.054207}
  (\bibinfo {year} {2024})\BibitemShut {NoStop}%
\bibitem [{\citenamefont {Kuno}\ \emph {et~al.}(2020)\citenamefont {Kuno},
  \citenamefont {Orito},\ and\ \citenamefont {Ichinose}}]{Kuno2020}%
  \BibitemOpen
  \bibfield  {author} {\bibinfo {author} {\bibfnamefont {Y.}~\bibnamefont
  {Kuno}}, \bibinfo {author} {\bibfnamefont {T.}~\bibnamefont {Orito}},\ and\
  \bibinfo {author} {\bibfnamefont {I.}~\bibnamefont {Ichinose}},\ }\bibfield
  {title} {\bibinfo {title} {Flat-band many-body localization and ergodicity
  breaking in the creutz ladder},\ }\href
  {https://doi.org/10.1088/1367-2630/ab6352} {\bibfield  {journal} {\bibinfo
  {journal} {New Journal of Physics}\ }\textbf {\bibinfo {volume} {22}},\
  \bibinfo {pages} {013032} (\bibinfo {year} {2020})}\BibitemShut {NoStop}%
\bibitem [{\citenamefont {Orito}\ \emph {et~al.}(2021)\citenamefont {Orito},
  \citenamefont {Kuno},\ and\ \citenamefont {Ichinose}}]{Orito2021}%
  \BibitemOpen
  \bibfield  {author} {\bibinfo {author} {\bibfnamefont {T.}~\bibnamefont
  {Orito}}, \bibinfo {author} {\bibfnamefont {Y.}~\bibnamefont {Kuno}},\ and\
  \bibinfo {author} {\bibfnamefont {I.}~\bibnamefont {Ichinose}},\ }\bibfield
  {title} {\bibinfo {title} {Nonthermalized dynamics of flat-band many-body
  localization},\ }\bibfield  {journal} {\bibinfo  {journal} {Physical Review
  B}\ }\textbf {\bibinfo {volume} {103}},\ \href
  {https://doi.org/10.1103/physrevb.103.l060301} {10.1103/physrevb.103.l060301}
  (\bibinfo {year} {2021})\BibitemShut {NoStop}%
\bibitem [{\citenamefont {Cheng}\ \emph {et~al.}(2024)\citenamefont {Cheng},
  \citenamefont {Yin},\ and\ \citenamefont {Yao}}]{Cheng2024}%
  \BibitemOpen
  \bibfield  {author} {\bibinfo {author} {\bibfnamefont {J.-Q.}\ \bibnamefont
  {Cheng}}, \bibinfo {author} {\bibfnamefont {S.}~\bibnamefont {Yin}},\ and\
  \bibinfo {author} {\bibfnamefont {D.-X.}\ \bibnamefont {Yao}},\ }\bibfield
  {title} {\bibinfo {title} {Dynamical localization transition in the
  non-hermitian lattice gauge theory},\ }\href
  {https://doi.org/10.1038/s42005-024-01544-6} {\bibfield  {journal} {\bibinfo
  {journal} {Communications Physics}\ }\textbf {\bibinfo {volume} {7}},\
  \bibinfo {pages} {58} (\bibinfo {year} {2024})}\BibitemShut {NoStop}%
\bibitem [{\citenamefont {Moudgalya}\ \emph {et~al.}(2022)\citenamefont
  {Moudgalya}, \citenamefont {Bernevig},\ and\ \citenamefont
  {Regnault}}]{Moudgalya2022}%
  \BibitemOpen
  \bibfield  {author} {\bibinfo {author} {\bibfnamefont {S.}~\bibnamefont
  {Moudgalya}}, \bibinfo {author} {\bibfnamefont {B.~A.}\ \bibnamefont
  {Bernevig}},\ and\ \bibinfo {author} {\bibfnamefont {N.}~\bibnamefont
  {Regnault}},\ }\bibfield  {title} {\bibinfo {title} {Quantum many-body scars
  and hilbert space fragmentation: a review of exact results},\ }\href
  {https://doi.org/10.1088/1361-6633/ac73a0} {\bibfield  {journal} {\bibinfo
  {journal} {Reports on Progress in Physics}\ }\textbf {\bibinfo {volume}
  {85}},\ \bibinfo {pages} {086501} (\bibinfo {year} {2022})}\BibitemShut
  {NoStop}%
\bibitem [{\citenamefont {Chandran}\ \emph {et~al.}(2023)\citenamefont
  {Chandran}, \citenamefont {Iadecola}, \citenamefont {Khemani},\ and\
  \citenamefont {Moessner}}]{Chandran2023}%
  \BibitemOpen
  \bibfield  {author} {\bibinfo {author} {\bibfnamefont {A.}~\bibnamefont
  {Chandran}}, \bibinfo {author} {\bibfnamefont {T.}~\bibnamefont {Iadecola}},
  \bibinfo {author} {\bibfnamefont {V.}~\bibnamefont {Khemani}},\ and\ \bibinfo
  {author} {\bibfnamefont {R.}~\bibnamefont {Moessner}},\ }\bibfield  {title}
  {\bibinfo {title} {Quantum many-body scars: A quasiparticle perspective},\
  }\href
  {https://doi.org/https://doi.org/10.1146/annurev-conmatphys-031620-101617}
  {\bibfield  {journal} {\bibinfo  {journal} {Annual Review of Condensed Matter
  Physics}\ }\textbf {\bibinfo {volume} {14}},\ \bibinfo {pages} {443}
  (\bibinfo {year} {2023})}\BibitemShut {NoStop}%
\bibitem [{\citenamefont {Ros}\ \emph {et~al.}(2015)\citenamefont {Ros},
  \citenamefont {Müller},\ and\ \citenamefont {Scardicchio}}]{Ros2015}%
  \BibitemOpen
  \bibfield  {author} {\bibinfo {author} {\bibfnamefont {V.}~\bibnamefont
  {Ros}}, \bibinfo {author} {\bibfnamefont {M.}~\bibnamefont {Müller}},\ and\
  \bibinfo {author} {\bibfnamefont {A.}~\bibnamefont {Scardicchio}},\
  }\bibfield  {title} {\bibinfo {title} {Integrals of motion in the many-body
  localized phase},\ }\href {https://doi.org/10.1016/j.nuclphysb.2014.12.014}
  {\bibfield  {journal} {\bibinfo  {journal} {Nuclear Physics B}\ }\textbf
  {\bibinfo {volume} {891}},\ \bibinfo {pages} {420–465} (\bibinfo {year}
  {2015})}\BibitemShut {NoStop}%
\bibitem [{\citenamefont {Serbyn}\ \emph {et~al.}(2013)\citenamefont {Serbyn},
  \citenamefont {Papić},\ and\ \citenamefont {Abanin}}]{Serbyn2013}%
  \BibitemOpen
  \bibfield  {author} {\bibinfo {author} {\bibfnamefont {M.}~\bibnamefont
  {Serbyn}}, \bibinfo {author} {\bibfnamefont {Z.}~\bibnamefont {Papić}},\
  and\ \bibinfo {author} {\bibfnamefont {D.~A.}\ \bibnamefont {Abanin}},\
  }\bibfield  {title} {\bibinfo {title} {Local conservation laws and the
  structure of the many-body localized states},\ }\bibfield  {journal}
  {\bibinfo  {journal} {Physical Review Letters}\ }\textbf {\bibinfo {volume}
  {111}},\ \href {https://doi.org/10.1103/physrevlett.111.127201}
  {10.1103/physrevlett.111.127201} (\bibinfo {year} {2013})\BibitemShut
  {NoStop}%
\bibitem [{\citenamefont {Bardarson}\ \emph {et~al.}(2012)\citenamefont
  {Bardarson}, \citenamefont {Pollmann},\ and\ \citenamefont
  {Moore}}]{Bardarson2012}%
  \BibitemOpen
  \bibfield  {author} {\bibinfo {author} {\bibfnamefont {J.~H.}\ \bibnamefont
  {Bardarson}}, \bibinfo {author} {\bibfnamefont {F.}~\bibnamefont
  {Pollmann}},\ and\ \bibinfo {author} {\bibfnamefont {J.~E.}\ \bibnamefont
  {Moore}},\ }\bibfield  {title} {\bibinfo {title} {Unbounded growth of
  entanglement in models of many-body localization},\ }\bibfield  {journal}
  {\bibinfo  {journal} {Physical Review Letters}\ }\textbf {\bibinfo {volume}
  {109}},\ \href {https://doi.org/10.1103/physrevlett.109.017202}
  {10.1103/physrevlett.109.017202} (\bibinfo {year} {2012})\BibitemShut
  {NoStop}%
\bibitem [{\citenamefont {Dumitrescu}\ \emph {et~al.}(2017)\citenamefont
  {Dumitrescu}, \citenamefont {Vasseur},\ and\ \citenamefont
  {Potter}}]{Dumitrescu2017}%
  \BibitemOpen
  \bibfield  {author} {\bibinfo {author} {\bibfnamefont {P.~T.}\ \bibnamefont
  {Dumitrescu}}, \bibinfo {author} {\bibfnamefont {R.}~\bibnamefont
  {Vasseur}},\ and\ \bibinfo {author} {\bibfnamefont {A.~C.}\ \bibnamefont
  {Potter}},\ }\bibfield  {title} {\bibinfo {title} {Scaling theory of
  entanglement at the many-body localization transition},\ }\bibfield
  {journal} {\bibinfo  {journal} {Physical Review Letters}\ }\textbf {\bibinfo
  {volume} {119}},\ \href {https://doi.org/10.1103/physrevlett.119.110604}
  {10.1103/physrevlett.119.110604} (\bibinfo {year} {2017})\BibitemShut
  {NoStop}%
\bibitem [{\citenamefont {GU}(2010)}]{GU2010}%
  \BibitemOpen
  \bibfield  {author} {\bibinfo {author} {\bibfnamefont {S.-J.}\ \bibnamefont
  {GU}},\ }\bibfield  {title} {\bibinfo {title} {Fidelity approach to quantum
  phase transitions},\ }\href {https://doi.org/10.1142/s0217979210056335}
  {\bibfield  {journal} {\bibinfo  {journal} {International Journal of Modern
  Physics B}\ }\textbf {\bibinfo {volume} {24}},\ \bibinfo {pages}
  {4371–4458} (\bibinfo {year} {2010})}\BibitemShut {NoStop}%
\bibitem [{\citenamefont {Sels}\ and\ \citenamefont
  {Polkovnikov}(2021)}]{Sels2021}%
  \BibitemOpen
  \bibfield  {author} {\bibinfo {author} {\bibfnamefont {D.}~\bibnamefont
  {Sels}}\ and\ \bibinfo {author} {\bibfnamefont {A.}~\bibnamefont
  {Polkovnikov}},\ }\bibfield  {title} {\bibinfo {title} {Dynamical obstruction
  to localization in a disordered spin chain},\ }\bibfield  {journal} {\bibinfo
   {journal} {Physical Review E}\ }\textbf {\bibinfo {volume} {104}},\ \href
  {https://doi.org/10.1103/physreve.104.054105} {10.1103/physreve.104.054105}
  (\bibinfo {year} {2021})\BibitemShut {NoStop}%
\bibitem [{\citenamefont {Kim}\ \emph {et~al.}(2025)\citenamefont {Kim},
  \citenamefont {Lim}, \citenamefont {Matirko}, \citenamefont {Polkovnikov},\
  and\ \citenamefont {Flynn}}]{kim2025definingclassicalquantumchaos}%
  \BibitemOpen
  \bibfield  {author} {\bibinfo {author} {\bibfnamefont {H.}~\bibnamefont
  {Kim}}, \bibinfo {author} {\bibfnamefont {C.}~\bibnamefont {Lim}}, \bibinfo
  {author} {\bibfnamefont {K.}~\bibnamefont {Matirko}}, \bibinfo {author}
  {\bibfnamefont {A.}~\bibnamefont {Polkovnikov}},\ and\ \bibinfo {author}
  {\bibfnamefont {M.~O.}\ \bibnamefont {Flynn}},\ }\bibfield  {title} {\bibinfo
  {title} {Defining classical and quantum chaos through adiabatic
  transformations},\ }\href {https://arxiv.org/abs/2401.01927} {\  (\bibinfo
  {year} {2025})},\ \Eprint {https://arxiv.org/abs/2401.01927}
  {arXiv:2401.01927 [cond-mat.stat-mech]} \BibitemShut {NoStop}%
\bibitem [{\citenamefont {Świętek}\ \emph {et~al.}(2025)\citenamefont
  {Świętek}, \citenamefont {Łydżba},\ and\ \citenamefont
  {Vidmar}}]{Swietek2025}%
  \BibitemOpen
  \bibfield  {author} {\bibinfo {author} {\bibfnamefont {R.}~\bibnamefont
  {Świętek}}, \bibinfo {author} {\bibfnamefont {P.}~\bibnamefont
  {Łydżba}},\ and\ \bibinfo {author} {\bibfnamefont {L.}~\bibnamefont
  {Vidmar}},\ }\bibfield  {title} {\bibinfo {title} {Fading ergodicity meets
  maximal chaos},\ }\href {https://doi.org/10.1103/PhysRevB.111.184203}
  {\bibfield  {journal} {\bibinfo  {journal} {Phys. Rev. B}\ }\textbf {\bibinfo
  {volume} {111}},\ \bibinfo {pages} {184203} (\bibinfo {year}
  {2025})}\BibitemShut {NoStop}%
\bibitem [{\citenamefont {Kolodrubetz}\ \emph {et~al.}(2017)\citenamefont
  {Kolodrubetz}, \citenamefont {Sels}, \citenamefont {Mehta},\ and\
  \citenamefont {Polkovnikov}}]{Kolodrubetz2017}%
  \BibitemOpen
  \bibfield  {author} {\bibinfo {author} {\bibfnamefont {M.}~\bibnamefont
  {Kolodrubetz}}, \bibinfo {author} {\bibfnamefont {D.}~\bibnamefont {Sels}},
  \bibinfo {author} {\bibfnamefont {P.}~\bibnamefont {Mehta}},\ and\ \bibinfo
  {author} {\bibfnamefont {A.}~\bibnamefont {Polkovnikov}},\ }\bibfield
  {title} {\bibinfo {title} {Geometry and non-adiabatic response in quantum and
  classical systems},\ }\href {https://doi.org/10.1016/j.physrep.2017.07.001}
  {\bibfield  {journal} {\bibinfo  {journal} {Physics Reports}\ }\textbf
  {\bibinfo {volume} {697}},\ \bibinfo {pages} {1–87} (\bibinfo {year}
  {2017})}\BibitemShut {NoStop}%
\bibitem [{\citenamefont {Pandey}\ \emph {et~al.}(2020)\citenamefont {Pandey},
  \citenamefont {Claeys}, \citenamefont {Campbell}, \citenamefont
  {Polkovnikov},\ and\ \citenamefont {Sels}}]{Pandey2020}%
  \BibitemOpen
  \bibfield  {author} {\bibinfo {author} {\bibfnamefont {M.}~\bibnamefont
  {Pandey}}, \bibinfo {author} {\bibfnamefont {P.~W.}\ \bibnamefont {Claeys}},
  \bibinfo {author} {\bibfnamefont {D.~K.}\ \bibnamefont {Campbell}}, \bibinfo
  {author} {\bibfnamefont {A.}~\bibnamefont {Polkovnikov}},\ and\ \bibinfo
  {author} {\bibfnamefont {D.}~\bibnamefont {Sels}},\ }\bibfield  {title}
  {\bibinfo {title} {Adiabatic eigenstate deformations as a sensitive probe for
  quantum chaos},\ }\bibfield  {journal} {\bibinfo  {journal} {Physical Review
  X}\ }\textbf {\bibinfo {volume} {10}},\ \href
  {https://doi.org/10.1103/physrevx.10.041017} {10.1103/physrevx.10.041017}
  (\bibinfo {year} {2020})\BibitemShut {NoStop}%
\bibitem [{\citenamefont {LeBlond}\ \emph {et~al.}(2021)\citenamefont
  {LeBlond}, \citenamefont {Sels}, \citenamefont {Polkovnikov},\ and\
  \citenamefont {Rigol}}]{LeBlond2021}%
  \BibitemOpen
  \bibfield  {author} {\bibinfo {author} {\bibfnamefont {T.}~\bibnamefont
  {LeBlond}}, \bibinfo {author} {\bibfnamefont {D.}~\bibnamefont {Sels}},
  \bibinfo {author} {\bibfnamefont {A.}~\bibnamefont {Polkovnikov}},\ and\
  \bibinfo {author} {\bibfnamefont {M.}~\bibnamefont {Rigol}},\ }\bibfield
  {title} {\bibinfo {title} {Universality in the onset of quantum chaos in
  many-body systems},\ }\bibfield  {journal} {\bibinfo  {journal} {Physical
  Review B}\ }\textbf {\bibinfo {volume} {104}},\ \href
  {https://doi.org/10.1103/physrevb.104.l201117} {10.1103/physrevb.104.l201117}
  (\bibinfo {year} {2021})\BibitemShut {NoStop}%
\bibitem [{\citenamefont {Kim}\ and\ \citenamefont
  {Polkovnikov}(2024)}]{Kim2024}%
  \BibitemOpen
  \bibfield  {author} {\bibinfo {author} {\bibfnamefont {H.}~\bibnamefont
  {Kim}}\ and\ \bibinfo {author} {\bibfnamefont {A.}~\bibnamefont
  {Polkovnikov}},\ }\bibfield  {title} {\bibinfo {title} {Integrability as an
  attractor of adiabatic flows},\ }\bibfield  {journal} {\bibinfo  {journal}
  {Physical Review B}\ }\textbf {\bibinfo {volume} {109}},\ \href
  {https://doi.org/10.1103/physrevb.109.195162} {10.1103/physrevb.109.195162}
  (\bibinfo {year} {2024})\BibitemShut {NoStop}%
\bibitem [{\citenamefont {Atas}\ \emph {et~al.}(2013)\citenamefont {Atas},
  \citenamefont {Bogomolny}, \citenamefont {Giraud},\ and\ \citenamefont
  {Roux}}]{Atas2013}%
  \BibitemOpen
  \bibfield  {author} {\bibinfo {author} {\bibfnamefont {Y.~Y.}\ \bibnamefont
  {Atas}}, \bibinfo {author} {\bibfnamefont {E.}~\bibnamefont {Bogomolny}},
  \bibinfo {author} {\bibfnamefont {O.}~\bibnamefont {Giraud}},\ and\ \bibinfo
  {author} {\bibfnamefont {G.}~\bibnamefont {Roux}},\ }\bibfield  {title}
  {\bibinfo {title} {Distribution of the ratio of consecutive level spacings in
  random matrix ensembles},\ }\bibfield  {journal} {\bibinfo  {journal}
  {Physical Review Letters}\ }\textbf {\bibinfo {volume} {110}},\ \href
  {https://doi.org/10.1103/physrevlett.110.084101}
  {10.1103/physrevlett.110.084101} (\bibinfo {year} {2013})\BibitemShut
  {NoStop}%
\bibitem [{\citenamefont {Santos}\ and\ \citenamefont
  {Rigol}(2010)}]{Santos2010}%
  \BibitemOpen
  \bibfield  {author} {\bibinfo {author} {\bibfnamefont {L.~F.}\ \bibnamefont
  {Santos}}\ and\ \bibinfo {author} {\bibfnamefont {M.}~\bibnamefont {Rigol}},\
  }\bibfield  {title} {\bibinfo {title} {Onset of quantum chaos in
  one-dimensional bosonic and fermionic systems and its relation to
  thermalization},\ }\bibfield  {journal} {\bibinfo  {journal} {Physical Review
  E}\ }\textbf {\bibinfo {volume} {81}},\ \href
  {https://doi.org/10.1103/physreve.81.036206} {10.1103/physreve.81.036206}
  (\bibinfo {year} {2010})\BibitemShut {NoStop}%
\bibitem [{\citenamefont {Giamarchi}(2003)}]{Giamarchi2003}%
  \BibitemOpen
  \bibfield  {author} {\bibinfo {author} {\bibfnamefont {T.}~\bibnamefont
  {Giamarchi}},\ }\href@noop {} {\emph {\bibinfo {title} {Quantum Physics in
  One Dimension}}}\ (\bibinfo  {publisher} {Oxford University Press},\ \bibinfo
  {year} {2003})\BibitemShut {NoStop}%
\bibitem [{\citenamefont {Jordan}\ and\ \citenamefont
  {Wigner}(1928)}]{Jordan1928}%
  \BibitemOpen
  \bibfield  {author} {\bibinfo {author} {\bibfnamefont {P.}~\bibnamefont
  {Jordan}}\ and\ \bibinfo {author} {\bibfnamefont {E.}~\bibnamefont
  {Wigner}},\ }\bibfield  {title} {\bibinfo {title} {{\"U}ber das paulische
  {\"a}quivalenzverbot},\ }\href {https://doi.org/10.1007/BF01331938}
  {\bibfield  {journal} {\bibinfo  {journal} {Z. Phys.}\ }\textbf {\bibinfo
  {volume} {47}},\ \bibinfo {pages} {631} (\bibinfo {year} {1928})}\BibitemShut
  {NoStop}%
\bibitem [{\citenamefont {Lieb}\ \emph {et~al.}(1961)\citenamefont {Lieb},
  \citenamefont {Schultz},\ and\ \citenamefont {Mattis}}]{Lieb1961}%
  \BibitemOpen
  \bibfield  {author} {\bibinfo {author} {\bibfnamefont {E.}~\bibnamefont
  {Lieb}}, \bibinfo {author} {\bibfnamefont {T.}~\bibnamefont {Schultz}},\ and\
  \bibinfo {author} {\bibfnamefont {D.}~\bibnamefont {Mattis}},\ }\bibfield
  {title} {\bibinfo {title} {Two soluble models of an antiferromagnetic
  chain},\ }\href {https://doi.org/10.1016/0003-4916(61)90115-4} {\bibfield
  {journal} {\bibinfo  {journal} {Ann. Phys.}\ }\textbf {\bibinfo {volume}
  {16}},\ \bibinfo {pages} {407} (\bibinfo {year} {1961})}\BibitemShut
  {NoStop}%
\bibitem [{\citenamefont {Anderson}(1958)}]{Anderson1958}%
  \BibitemOpen
  \bibfield  {author} {\bibinfo {author} {\bibfnamefont {P.~W.}\ \bibnamefont
  {Anderson}},\ }\bibfield  {title} {\bibinfo {title} {Absence of diffusion in
  certain random lattices},\ }\href {https://doi.org/10.1103/PhysRev.109.1492}
  {\bibfield  {journal} {\bibinfo  {journal} {Phys. Rev.}\ }\textbf {\bibinfo
  {volume} {109}},\ \bibinfo {pages} {1492} (\bibinfo {year}
  {1958})}\BibitemShut {NoStop}%
\bibitem [{\citenamefont {Khemani}\ \emph {et~al.}(2017)\citenamefont
  {Khemani}, \citenamefont {Lim}, \citenamefont {Sheng},\ and\ \citenamefont
  {Huse}}]{Khemani2017}%
  \BibitemOpen
  \bibfield  {author} {\bibinfo {author} {\bibfnamefont {V.}~\bibnamefont
  {Khemani}}, \bibinfo {author} {\bibfnamefont {S.}~\bibnamefont {Lim}},
  \bibinfo {author} {\bibfnamefont {D.}~\bibnamefont {Sheng}},\ and\ \bibinfo
  {author} {\bibfnamefont {D.~A.}\ \bibnamefont {Huse}},\ }\bibfield  {title}
  {\bibinfo {title} {Critical properties of the many-body localization
  transition},\ }\bibfield  {journal} {\bibinfo  {journal} {Physical Review X}\
  }\textbf {\bibinfo {volume} {7}},\ \href
  {https://doi.org/10.1103/physrevx.7.021013} {10.1103/physrevx.7.021013}
  (\bibinfo {year} {2017})\BibitemShut {NoStop}%
\bibitem [{\citenamefont {Amico}\ \emph {et~al.}(2008)\citenamefont {Amico},
  \citenamefont {Fazio}, \citenamefont {Osterloh},\ and\ \citenamefont
  {Vedral}}]{Amico2008}%
  \BibitemOpen
  \bibfield  {author} {\bibinfo {author} {\bibfnamefont {L.}~\bibnamefont
  {Amico}}, \bibinfo {author} {\bibfnamefont {R.}~\bibnamefont {Fazio}},
  \bibinfo {author} {\bibfnamefont {A.}~\bibnamefont {Osterloh}},\ and\
  \bibinfo {author} {\bibfnamefont {V.}~\bibnamefont {Vedral}},\ }\bibfield
  {title} {\bibinfo {title} {Entanglement in many-body systems},\ }\href
  {https://doi.org/10.1103/RevModPhys.80.517} {\bibfield  {journal} {\bibinfo
  {journal} {Rev. Mod. Phys.}\ }\textbf {\bibinfo {volume} {80}},\ \bibinfo
  {pages} {517} (\bibinfo {year} {2008})}\BibitemShut {NoStop}%
\bibitem [{\citenamefont {Laflorencie}(2016)}]{Laflorencie2016}%
  \BibitemOpen
  \bibfield  {author} {\bibinfo {author} {\bibfnamefont {N.}~\bibnamefont
  {Laflorencie}},\ }\bibfield  {title} {\bibinfo {title} {Quantum entanglement
  in condensed matter systems},\ }\href
  {https://doi.org/10.1016/j.physrep.2016.06.008} {\bibfield  {journal}
  {\bibinfo  {journal} {Physics Reports}\ }\textbf {\bibinfo {volume} {646}},\
  \bibinfo {pages} {1–59} (\bibinfo {year} {2016})}\BibitemShut {NoStop}%
\bibitem [{\citenamefont {You}\ \emph {et~al.}(2007)\citenamefont {You},
  \citenamefont {Li},\ and\ \citenamefont {Gu}}]{You2007}%
  \BibitemOpen
  \bibfield  {author} {\bibinfo {author} {\bibfnamefont {W.-L.}\ \bibnamefont
  {You}}, \bibinfo {author} {\bibfnamefont {Y.-W.}\ \bibnamefont {Li}},\ and\
  \bibinfo {author} {\bibfnamefont {S.-J.}\ \bibnamefont {Gu}},\ }\bibfield
  {title} {\bibinfo {title} {Fidelity, dynamic structure factor, and
  susceptibility in critical phenomena},\ }\bibfield  {journal} {\bibinfo
  {journal} {Physical Review E}\ }\textbf {\bibinfo {volume} {76}},\ \href
  {https://doi.org/10.1103/physreve.76.022101} {10.1103/physreve.76.022101}
  (\bibinfo {year} {2007})\BibitemShut {NoStop}%
\bibitem [{\citenamefont {Sels}\ and\ \citenamefont
  {Polkovnikov}(2023)}]{Sels2023}%
  \BibitemOpen
  \bibfield  {author} {\bibinfo {author} {\bibfnamefont {D.}~\bibnamefont
  {Sels}}\ and\ \bibinfo {author} {\bibfnamefont {A.}~\bibnamefont
  {Polkovnikov}},\ }\bibfield  {title} {\bibinfo {title} {Thermalization of
  dilute impurities in one-dimensional spin chains},\ }\href
  {https://doi.org/10.1103/PhysRevX.13.011041} {\bibfield  {journal} {\bibinfo
  {journal} {Phys. Rev. X}\ }\textbf {\bibinfo {volume} {13}},\ \bibinfo
  {pages} {011041} (\bibinfo {year} {2023})}\BibitemShut {NoStop}%
\bibitem [{\citenamefont {Sels}\ and\ \citenamefont
  {Polkovnikov}(2017)}]{Sels_2017}%
  \BibitemOpen
  \bibfield  {author} {\bibinfo {author} {\bibfnamefont {D.}~\bibnamefont
  {Sels}}\ and\ \bibinfo {author} {\bibfnamefont {A.}~\bibnamefont
  {Polkovnikov}},\ }\bibfield  {title} {\bibinfo {title} {Minimizing
  irreversible losses in quantum systems by local counterdiabatic driving},\
  }\bibfield  {journal} {\bibinfo  {journal} {Proceedings of the National
  Academy of Sciences}\ }\textbf {\bibinfo {volume} {114}},\ \href
  {https://doi.org/10.1073/pnas.1619826114} {10.1073/pnas.1619826114} (\bibinfo
  {year} {2017})\BibitemShut {NoStop}%
\bibitem [{\citenamefont {Bianchi}\ \emph {et~al.}(2022)\citenamefont
  {Bianchi}, \citenamefont {Hackl}, \citenamefont {Kieburg}, \citenamefont
  {Rigol},\ and\ \citenamefont {Vidmar}}]{Bianchi2022}%
  \BibitemOpen
  \bibfield  {author} {\bibinfo {author} {\bibfnamefont {E.}~\bibnamefont
  {Bianchi}}, \bibinfo {author} {\bibfnamefont {L.}~\bibnamefont {Hackl}},
  \bibinfo {author} {\bibfnamefont {M.}~\bibnamefont {Kieburg}}, \bibinfo
  {author} {\bibfnamefont {M.}~\bibnamefont {Rigol}},\ and\ \bibinfo {author}
  {\bibfnamefont {L.}~\bibnamefont {Vidmar}},\ }\bibfield  {title} {\bibinfo
  {title} {Volume-law entanglement entropy of typical pure quantum states},\
  }\bibfield  {journal} {\bibinfo  {journal} {PRX Quantum}\ }\textbf {\bibinfo
  {volume} {3}},\ \href {https://doi.org/10.1103/prxquantum.3.030201}
  {10.1103/prxquantum.3.030201} (\bibinfo {year} {2022})\BibitemShut {NoStop}%
\bibitem [{\citenamefont {Ben-Zion}\ \emph {et~al.}(2020)\citenamefont
  {Ben-Zion}, \citenamefont {McGreevy},\ and\ \citenamefont
  {Grover}}]{Ben_Zion_2020}%
  \BibitemOpen
  \bibfield  {author} {\bibinfo {author} {\bibfnamefont {D.}~\bibnamefont
  {Ben-Zion}}, \bibinfo {author} {\bibfnamefont {J.}~\bibnamefont {McGreevy}},\
  and\ \bibinfo {author} {\bibfnamefont {T.}~\bibnamefont {Grover}},\
  }\bibfield  {title} {\bibinfo {title} {Disentangling quantum matter with
  measurements},\ }\bibfield  {journal} {\bibinfo  {journal} {Physical Review
  B}\ }\textbf {\bibinfo {volume} {101}},\ \href
  {https://doi.org/10.1103/physrevb.101.115131} {10.1103/physrevb.101.115131}
  (\bibinfo {year} {2020})\BibitemShut {NoStop}%
\bibitem [{\citenamefont {Bloch}(1958)}]{BLOCH1958329}%
  \BibitemOpen
  \bibfield  {author} {\bibinfo {author} {\bibfnamefont {C.}~\bibnamefont
  {Bloch}},\ }\bibfield  {title} {\bibinfo {title} {Sur la théorie des
  perturbations des états liés},\ }\href
  {https://doi.org/https://doi.org/10.1016/0029-5582(58)90116-0} {\bibfield
  {journal} {\bibinfo  {journal} {Nuclear Physics}\ }\textbf {\bibinfo {volume}
  {6}},\ \bibinfo {pages} {329} (\bibinfo {year} {1958})}\BibitemShut {NoStop}%
\bibitem [{\citenamefont {Winkler}(2003)}]{Winkler2003}%
  \BibitemOpen
  \bibfield  {author} {\bibinfo {author} {\bibfnamefont {R.}~\bibnamefont
  {Winkler}},\ }\href {https://doi.org/10.1007/b13586} {\emph {\bibinfo {title}
  {Spin-Orbit Coupling Effects in Two-Dimensional Electron and Hole
  Systems}}},\ \bibinfo {edition} {1st}\ ed.,\ \bibinfo {series} {Springer
  Tracts in Modern Physics}, Vol.\ \bibinfo {volume} {191}\ (\bibinfo
  {publisher} {Springer},\ \bibinfo {address} {Berlin, Heidelberg},\ \bibinfo
  {year} {2003})\ pp.\ \bibinfo {pages} {XII, 228}\BibitemShut {NoStop}%
\bibitem [{\citenamefont {Bravyi}\ \emph {et~al.}(2011)\citenamefont {Bravyi},
  \citenamefont {DiVincenzo},\ and\ \citenamefont {Loss}}]{Bravyi2011}%
  \BibitemOpen
  \bibfield  {author} {\bibinfo {author} {\bibfnamefont {S.}~\bibnamefont
  {Bravyi}}, \bibinfo {author} {\bibfnamefont {D.~P.}\ \bibnamefont
  {DiVincenzo}},\ and\ \bibinfo {author} {\bibfnamefont {D.}~\bibnamefont
  {Loss}},\ }\bibfield  {title} {\bibinfo {title} {Schrieffer–wolff
  transformation for quantum many-body systems},\ }\href
  {https://doi.org/10.1016/j.aop.2011.06.004} {\bibfield  {journal} {\bibinfo
  {journal} {Annals of Physics}\ }\textbf {\bibinfo {volume} {326}},\ \bibinfo
  {pages} {2793–2826} (\bibinfo {year} {2011})}\BibitemShut {NoStop}%
\bibitem [{\citenamefont {MacDonald}\ \emph {et~al.}(1988)\citenamefont
  {MacDonald}, \citenamefont {Girvin},\ and\ \citenamefont
  {Yoshioka}}]{MacDonald1988}%
  \BibitemOpen
  \bibfield  {author} {\bibinfo {author} {\bibfnamefont {A.~H.}\ \bibnamefont
  {MacDonald}}, \bibinfo {author} {\bibfnamefont {S.~M.}\ \bibnamefont
  {Girvin}},\ and\ \bibinfo {author} {\bibfnamefont {D.}~\bibnamefont
  {Yoshioka}},\ }\bibfield  {title} {\bibinfo {title} {$\frac{t}{U}$ expansion
  for the hubbard model},\ }\href {https://doi.org/10.1103/PhysRevB.37.9753}
  {\bibfield  {journal} {\bibinfo  {journal} {Phys. Rev. B}\ }\textbf {\bibinfo
  {volume} {37}},\ \bibinfo {pages} {9753} (\bibinfo {year}
  {1988})}\BibitemShut {NoStop}%
\bibitem [{\citenamefont {Cohen-Tannoudji}\ \emph {et~al.}(1992)\citenamefont
  {Cohen-Tannoudji}, \citenamefont {Dupont-Roc},\ and\ \citenamefont
  {Grynberg}}]{CohenTannoudji1992}%
  \BibitemOpen
  \bibfield  {author} {\bibinfo {author} {\bibfnamefont {C.}~\bibnamefont
  {Cohen-Tannoudji}}, \bibinfo {author} {\bibfnamefont {J.}~\bibnamefont
  {Dupont-Roc}},\ and\ \bibinfo {author} {\bibfnamefont {G.}~\bibnamefont
  {Grynberg}},\ }\href@noop {} {\emph {\bibinfo {title} {Atom--Photon
  Interactions: Basic Processes and Applications}}}\ (\bibinfo  {publisher}
  {Wiley},\ \bibinfo {year} {1992})\BibitemShut {NoStop}%
\bibitem [{\citenamefont {Hubac}\ and\ \citenamefont
  {Wilson}(2000)}]{Hubac2000}%
  \BibitemOpen
  \bibfield  {author} {\bibinfo {author} {\bibfnamefont {I.}~\bibnamefont
  {Hubac}}\ and\ \bibinfo {author} {\bibfnamefont {S.}~\bibnamefont {Wilson}},\
  }\bibfield  {title} {\bibinfo {title} {On the use of brillouin-wigner
  perturbation theory for many-body systems},\ }\href
  {https://doi.org/10.1088/0953-4075/33/3/306} {\bibfield  {journal} {\bibinfo
  {journal} {J. Phys. B: At. Mol. Opt. Phys.}\ }\textbf {\bibinfo {volume}
  {33}},\ \bibinfo {pages} {365} (\bibinfo {year} {2000})}\BibitemShut
  {NoStop}%
\bibitem [{\citenamefont {Papić}\ \emph {et~al.}(2015)\citenamefont {Papić},
  \citenamefont {Stoudenmire},\ and\ \citenamefont {Abanin}}]{Papi2015}%
  \BibitemOpen
  \bibfield  {author} {\bibinfo {author} {\bibfnamefont {Z.}~\bibnamefont
  {Papić}}, \bibinfo {author} {\bibfnamefont {E.~M.}\ \bibnamefont
  {Stoudenmire}},\ and\ \bibinfo {author} {\bibfnamefont {D.~A.}\ \bibnamefont
  {Abanin}},\ }\bibfield  {title} {\bibinfo {title} {Many-body localization in
  disorder-free systems: The importance of finite-size constraints},\ }\href
  {https://doi.org/10.1016/j.aop.2015.08.024} {\bibfield  {journal} {\bibinfo
  {journal} {Annals of Physics}\ }\textbf {\bibinfo {volume} {362}},\ \bibinfo
  {pages} {714–725} (\bibinfo {year} {2015})}\BibitemShut {NoStop}%
\bibitem [{\citenamefont {Weinberg}\ and\ \citenamefont
  {Bukov}(2017)}]{QuSpin2017}%
  \BibitemOpen
  \bibfield  {author} {\bibinfo {author} {\bibfnamefont {P.}~\bibnamefont
  {Weinberg}}\ and\ \bibinfo {author} {\bibfnamefont {M.}~\bibnamefont
  {Bukov}},\ }\bibfield  {title} {\bibinfo {title} {{QuSpin: a Python package
  for dynamics and exact diagonalisation of quantum many body systems part I:
  spin chains}},\ }\href {https://doi.org/10.21468/SciPostPhys.2.1.003}
  {\bibfield  {journal} {\bibinfo  {journal} {SciPost Phys.}\ }\textbf
  {\bibinfo {volume} {2}},\ \bibinfo {pages} {003} (\bibinfo {year}
  {2017})}\BibitemShut {NoStop}%
\bibitem [{\citenamefont {Weinberg}\ and\ \citenamefont
  {Bukov}(2019)}]{QuSpin2019}%
  \BibitemOpen
  \bibfield  {author} {\bibinfo {author} {\bibfnamefont {P.}~\bibnamefont
  {Weinberg}}\ and\ \bibinfo {author} {\bibfnamefont {M.}~\bibnamefont
  {Bukov}},\ }\bibfield  {title} {\bibinfo {title} {{QuSpin: a Python package
  for dynamics and exact diagonalisation of quantum many body systems. Part II:
  bosons, fermions and higher spins}},\ }\href
  {https://doi.org/10.21468/SciPostPhys.7.2.020} {\bibfield  {journal}
  {\bibinfo  {journal} {SciPost Phys.}\ }\textbf {\bibinfo {volume} {7}},\
  \bibinfo {pages} {020} (\bibinfo {year} {2019})}\BibitemShut {NoStop}%
\bibitem [{\citenamefont {Akaike}(1974)}]{Akaike1974}%
  \BibitemOpen
  \bibfield  {author} {\bibinfo {author} {\bibfnamefont {H.}~\bibnamefont
  {Akaike}},\ }\bibfield  {title} {\bibinfo {title} {A new look at the
  statistical model identification},\ }\href
  {https://doi.org/10.1109/TAC.1974.1100705} {\bibfield  {journal} {\bibinfo
  {journal} {IEEE Transactions on Automatic Control}\ }\textbf {\bibinfo
  {volume} {19}},\ \bibinfo {pages} {716} (\bibinfo {year} {1974})}\BibitemShut
  {NoStop}%
\bibitem [{\citenamefont {HURVICH}\ and\ \citenamefont
  {TSAI}(1989)}]{Hurvich1989}%
  \BibitemOpen
  \bibfield  {author} {\bibinfo {author} {\bibfnamefont {C.~M.}\ \bibnamefont
  {HURVICH}}\ and\ \bibinfo {author} {\bibfnamefont {C.-L.}\ \bibnamefont
  {TSAI}},\ }\bibfield  {title} {\bibinfo {title} {Regression and time series
  model selection in small samples},\ }\href
  {https://doi.org/10.1093/biomet/76.2.297} {\bibfield  {journal} {\bibinfo
  {journal} {Biometrika}\ }\textbf {\bibinfo {volume} {76}},\ \bibinfo {pages}
  {297} (\bibinfo {year} {1989})},\ \Eprint
  {https://arxiv.org/abs/https://academic.oup.com/biomet/article-pdf/76/2/297/737009/76-2-297.pdf}
  {https://academic.oup.com/biomet/article-pdf/76/2/297/737009/76-2-297.pdf}
  \BibitemShut {NoStop}%
\bibitem [{\citenamefont {Burnham}\ and\ \citenamefont
  {Anderson}(2002)}]{Burnham2002}%
  \BibitemOpen
  \bibfield  {author} {\bibinfo {author} {\bibfnamefont {K.~P.}\ \bibnamefont
  {Burnham}}\ and\ \bibinfo {author} {\bibfnamefont {D.~R.}\ \bibnamefont
  {Anderson}},\ }\href {https://doi.org/10.1007/b97636} {\emph {\bibinfo
  {title} {Model Selection and Multimodel Inference: A Practical
  Information-Theoretic Approach}}},\ \bibinfo {edition} {2nd}\ ed.\ (\bibinfo
  {publisher} {Springer},\ \bibinfo {address} {New York, NY},\ \bibinfo {year}
  {2002})\ \bibinfo {note} {hardcover ISBN 978-0-387-95364-9 (2002); Softcover
  ISBN 978-1-4419-2973-0 (2010); eBook ISBN 978-0-387-22456-5
  (2007)}\BibitemShut {NoStop}%
\bibitem [{\citenamefont {Stone}(1977)}]{Stone1977}%
  \BibitemOpen
  \bibfield  {author} {\bibinfo {author} {\bibfnamefont {M.}~\bibnamefont
  {Stone}},\ }\bibfield  {title} {\bibinfo {title} {An asymptotic equivalence
  of choice of model by cross-validation and akaike's criterion},\ }\href
  {http://www.jstor.org/stable/2984877} {\bibfield  {journal} {\bibinfo
  {journal} {Journal of the Royal Statistical Society. Series B
  (Methodological)}\ }\textbf {\bibinfo {volume} {39}},\ \bibinfo {pages} {44}
  (\bibinfo {year} {1977})}\BibitemShut {NoStop}%
\bibitem [{\citenamefont {Schwarz}(1978)}]{Schwarz1978}%
  \BibitemOpen
  \bibfield  {author} {\bibinfo {author} {\bibfnamefont {G.}~\bibnamefont
  {Schwarz}},\ }\bibfield  {title} {\bibinfo {title} {Estimating the dimension
  of a model},\ }\href {http://www.jstor.org/stable/2958889} {\bibfield
  {journal} {\bibinfo  {journal} {The Annals of Statistics}\ }\textbf {\bibinfo
  {volume} {6}},\ \bibinfo {pages} {461} (\bibinfo {year} {1978})}\BibitemShut
  {NoStop}%
\end{thebibliography}

%

\end{document}